\documentclass[a4paper,11pt]{article}
\pdfoutput=1
\usepackage{jheppub}
\usepackage{multirow}
\usepackage{array,hhline}
\usepackage{makecell}
\usepackage{hyperref}
\usepackage{booktabs}
\usepackage{floatrow}
\usepackage{autobreak}
\usepackage{longtable}
\usepackage{caption}
\usepackage{bbding}

\usepackage{bbm}

\allowdisplaybreaks[1]

\newcommand\scalemath[2]{\scalebox{#1}{\mbox{\ensuremath{\displaystyle #2}}}}
\newcommand{\nn}{\nonumber \\}
\newcommand{\Ax}{\ensuremath \mathbb{A}_x}
\newcommand{\Ay}{\ensuremath \mathbb{A}_y}
\newcommand{\Az}{\ensuremath \mathbb{A}_z}

\newcommand{\SKLP}{State Key Laboratory of Particle Detection and Electronics, University of Science and Technology of China, Hefei 230026, Anhui, People’s Republic of China}
\newcommand{\USTC}{Department of Modern Physics, University of Science and Technology of China, Hefei 230026, Anhui, People’s Republic of China}

\newcommand{\minus}{\ensuremath \scalebox{0.5}[1.0]{\( - \)}}
\newcommand{\pminus}{\hphantom{\minus}}

\author[a,b]{Ming-Ming Long,}
\author[a,b]{Ren-You Zhang,}
\author[a,b]{Wen-Gan Ma,}
\author[a,b]{Yi Jiang,}
\author[a,b]{Liang Han,}
\author[a,b]{Zhe Li,}
\author[a,b]{Shuai-Shuai Wang}

\affiliation[a]{\SKLP}
\affiliation[b]{\USTC}

\emailAdd{heplmm@mail.ustc.edu.cn}
\emailAdd{zhangry@ustc.edu.cn}
\emailAdd{mawg@ustc.edu.cn}
\emailAdd{jiangyi@ustc.edu.cn}
\emailAdd{hanl@ustc.edu.cn}
\emailAdd{brucelee@mail.ustc.edu.cn}
\emailAdd{wang1996@mail.ustc.edu.cn}

\title{Master integrals for mixed QCD-QED corrections to charged-current Drell-Yan production of a massive charged lepton}

\keywords{Master Integrals, Differential Equations, Drell-Yan production}

\abstract{
The master integrals for the mixed QCD-QED two-loop virtual corrections to the charged-current Drell-Yan process $q\bar{q}^{\prime} \rightarrow \ell \nu$ are computed analytically by using the differential equation method. A suitable choice of master integrals makes it successful to cast the differential equation system into the canonical form. We keep the dependence on charged lepton mass in the building of differential equations and then expand the system against the ratio of small charged lepton mass to large $W$-boson mass. In such a way the final results will contain large logarithms of the form $\log(m_{\ell}^2/m_W^2)$. Finally, all the canonical master integrals are given as Taylor series around $d = 4$ spacetime dimensions up to order four, with coefficients expressed in terms of Goncharov polylogarithms up to weight four.}

\begin{document}

\maketitle
\flushbottom

\section{Introduction}
\label{sec:1}
\par
With decades of accumulation of diligent investigations from both experimental and theoretical communities, the Drell-Yan (DY) processes \cite{Drell:1970wh} have become one of the best understood physics subjects at the LHC. They provide an excellent environment to probe the inner structure of proton \cite{Aaboud:2016btc} and the nature of the electroweak (EW) sector of the Standard Model (SM) by precision measurements of many EW observables, such as $W$-boson mass and width \cite{Aaboud:2017svj,Camarda:2016twt}, Weinberg weak mixing angle \cite{Aad:2015uau,Sirunyan:2018swq} and charge asymmetries \cite{Khachatryan:2016pev, Sirunyan:2020oum}. These are built on the huge amount of event data collected at colliders thanks to the large cross sections and clean experimental signatures of DY processes. In addition to the precision test of the SM, DY processes could help to search for signals of new physics as well because the new vector bosons predicted by conjectured extensions of the SM can be produced by a similar mechanism \cite{Aaboud:2016zkn,Khachatryan:2016jww, Aaboud:2017efa,Aad:2019wvl,Aaboud:2016cth,Khachatryan:2016zqb,Aad:2019fac}. Precision measurements of DY processes must be confronted with equally precise theoretical predictions, therefore the theoretical perturbative descriptions of DY processes should be extended to higher orders.

\par
At the parton level, the DY processes at the lowest-order can be categorized into neutral-current processes ($q\bar{q} \rightarrow \gamma/Z \rightarrow \ell^+\ell^-$) and charged-current processes ($q\bar{q}^{\prime} \rightarrow W \rightarrow \ell \nu$). The production rates of those DY processes were computed up to the QCD next-to-next-leading order (NNLO) decades ago \cite{Altarelli:1979ub,Matsuura:1988sm,Hamberg:1990np,Harlander:2002wh}, which turned out to be a great triumph in precision study of phenomenology of particle physics. In Refs.\cite{Anastasiou:2003ds,Melnikov:2006di,Melnikov:2006kv,Catani:2009sm,Catani:2010en}, $W$ and $Z$ productions in hadron collisions were studied fully differentially at NNLO in QCD. The threshold corrections to DY production at $\text{N}^3\text{LO}$ in QCD were presented in Refs.\cite{Ahmed:2014cla,Ahmed:2014uya}. Recently, progress has been made toward $\text{N}^3\text{LO}$ at the inclusive level for both neutral- and charged-current DY processes \cite{Duhr:2020seh,Duhr:2020sdp}, and the lepton pair rapidity distribution in the photon-mediated DY process at $\text{N}^3\text{LO}$ in $\alpha_s$ was calculated in Ref.\cite{Chen:2021vtu} for the first time. Due to the leptonic final state of DY processes, the QCD corrections are only involved in the initial state. However, the EW radiative corrections are more complicated; they cannot be fully factorized into the corrections to the production and decay of the intermediate gauge boson separately. The NLO EW corrections to the neutral- and charged-current DY processes have been studied in Refs.\cite{Baur:1997wa,Baur:2001ze,Zykunov:2006yb,Zykunov:2005tc,CarloniCalame:2007cd,Arbuzov:2007db,Dittmaier:2009cr} and \cite{Wackeroth:1996hz,Baur:1998kt,Dittmaier:2001ay,Baur:2004ig, Arbuzov:2005dd,CarloniCalame:2006zq,Brensing:2007qm}, respectively. A fully differential description of NLO corrections to DY processes has been implemented in automated Monte Carlo (MC) programs. For a review on QCD and EW radiative corrections to $W$- and $Z$-boson observables, see Ref.\cite{Alioli:2016fum}. The authors presented a systematic comparison of a dozen MC codes, which served as either a general MC framework or a dedicated program for (partial) DY processes. There are also some implementations at QCD NNLO, such as \texttt{GENEVA} \cite{Alioli:2015toa, Alioli:2016wqt}, \texttt{DYTurbo} (improved reimplementation of \texttt{DYqT}, \texttt{DYRes}, \texttt{DYNNLO}) \cite{Camarda:2019zyx}, \texttt{MCFM} \cite{Boughezal:2016wmq}, \texttt{MATRIX} \cite{Grazzini:2017mhc} and $\texttt{MiNNLO}_\texttt{PS}$ \cite{Monni:2019whf}. A comparison of those NNLO MC generators was provided in Ref.\cite{Alekhin:2021xcu} very recently.

\par
To build an exhaustive comprehension of DY processes at NNLO, it is necessary to consider the mixed QCD-EW corrections which are not yet available. The mixed QCD-EW NNLO corrections to DY processes can be divided into two categories: factorizable and non-factorizable. For $Z$-boson production in hadron collisions, the mixed QCD-QED corrections have been studied at both inclusive and exclusive levels in Refs.\cite{deFlorian:2018wcj,Delto:2019ewv,Cieri:2020ikq}, and the mixed QCD-EW corrections are given in Refs.\cite{Bonciani:2019nuy, Buccioni:2020cfi,Bonciani:2020tvf}. As for $W$-boson production, the mixed QCD-EW corrections can be found in Ref.\cite{Behring:2020cqi}. The radiative corrections of $\mathcal{O}(N_f\alpha_s\alpha)$ to off-shell $W/Z$ production at the LHC are also provided in Ref.\cite{Dittmaier:2020vra}. The non-factorizable initial-final corrections to DY processes in the resonance region can be estimated via the so-called pole approximation \cite{Dittmaier:2014qza,Dittmaier:2015rxo}. In the precision calculation of the non-factorizable mixed QCD-EW corrections, the primary technical problem is the evaluation of massive two-loop four-point master integrals (MIs) induced by the most complicated box-type Feynman diagrams. The two-loop box-type MIs have been calculated analytically and numerically in the approximation of $m_W = m_Z$ and $m_{\ell} = 0$ \cite{Bonciani:2016ypc,vonManteuffel:2017myy,Heller:2019gkq}. Recently, the mixed QCD-EW two-loop scattering amplitude for $q\bar{q} \rightarrow \ell^+\ell^-~ (m_{\ell} = 0)$ in dimensional regularization and its independence on $\gamma^5$ scheme were studied in detail in Ref.\cite{Heller:2020owb}, and the complete mixed QCD-EW NNLO corrections to the DY production of a massless lepton pair were first computed in Ref.\cite{Bonciani:2021zzf}.

\par
When lepton mass is taken into consideration, the collinear photon emission off the final-state charged lepton(s) gives rise to the radiative corrections that are enhanced by large logarithms of the form $\log(m_{\ell}^2/Q^2)$, where $Q$ stands for a characteristic scale of DY processes, like weak gauge boson mass or center-of-mass colliding energy. As is well known, those logarithms cancel exactly if collinear lepton-photon systems are treated fully inclusively, which is guaranteed by the Kinoshita-Lee-Nauenberg theorem \cite{Kinoshita:1962ur,Lee:1964is}. In the presence of phase-space cuts and in kinematic distributions, in general, the mass-singular contributions survive, but the collinear safety can be restored after performing the photon recombination procedure \cite{Baur:1998kt,Dittmaier:2001ay,Baur:2004ig,Brensing:2007qm}. Those mass-singular contributions can be fully captured by considering only the QED part of the EW corrections, since they are only induced by collinear photon radiation from charged leptons. The analytic calculation of the MIs for the mixed QCD-QED corrections to DY production of a massive lepton pair has been accomplished \cite{Hasan:2020vwn}. In this paper we focus on the charged-current DY process $q\bar{q}^{\prime} \rightarrow \ell \nu$ with non-zero lepton mass, and calculate the MIs for the mixed QCD-QED two-loop virtual corrections.

\par
The rest of this paper is organized as follows. In Section \ref{sec:2} the notation and conventions are declared. In Section \ref{sec:3}, the canonical differential equations are constructed, and the $46$ canonical MIs for the mixed QCD-QED two-loop virtual corrections to $q\bar{q}^{\prime} \rightarrow \ell \nu$ are evaluated analytically. Finally, a short summary is given in Section \ref{sec:4}.

\section{Notation and conventions}
\label{sec:2}
\par
In this paper we study the charged-current DY process
\begin{equation}
q(p_1) + \bar{q}^{\prime}(p_2) \rightarrow \ell(p_3) + \nu(p_4)\,.
\end{equation}
The light quarks are considered as massless while the charged lepton mass is kept non-zero throughout our calculation. All the external particles are on their mass-shell, i.e., $p_1^2 = p_2^2 = p_4^2 = 0$ and $p_3^2 = m_{\ell}^2$, and the scattering amplitude can be expressed in terms of the Mandelstam invariants
\begin{equation}
s=(p_1+p_2)^2\,,
\qquad
t=(p_2-p_3)^2\,,
\qquad
u=(p_1-p_3)^2\,,
\end{equation}
which satisfy $s + t + u = m_{\ell}^2$ due to momentum conservation. The mixed QCD-QED two-loop virtual corrections to $q\bar{q}^{\prime} \rightarrow \ell \nu$ can be classified as follows:
\begin{itemize}
\item {\it Non-factorizable corrections}

      All the non-factorizable mixed QCD-QED two-loop Feynman diagrams are box type and irreducible. The dimensionally regularized two-loop four-point Feynman integrals in $d = 4 - 2 \epsilon$ dimensions have the form as
      \begin{equation}
      F(n_1, ..., n_9)
      =
      \int \mathcal{D}^d l_1 \mathcal{D}^d l_2
      \frac{1}{D_1^{n_1} ... D_9^{n_9}}\,,
      \qquad
      (n_i \in \mathbb{Z},\quad i = 1, ..., 9)\,,
      \label{eq:FI}
      \end{equation}
      where $l_{1,2}$ are loop momenta and the integration measure is defined by
      \begin{equation}
      \mathcal{D}^d l_i
      =
      \frac{d^dl_i}{(2\pi)^d}
      \left(
      \frac{i S_{\epsilon}}{16\pi^2}
      \right)^{-1}
      \quad
      \text{with}
      \quad
      S_{\epsilon}
      =
      (4\pi)^{\epsilon}
      \frac{\Gamma(1+\epsilon)\Gamma(1-\epsilon)^2}{\Gamma(1-2\epsilon)}\,.
      \label{eq:measure}
      \end{equation}
      For arbitrary $n_i \in \mathbb{Z}$, $i=1, ..., 9$, the set of all such $F(n_1, ..., n_9)$ in Eq.(\ref{eq:FI}) is called an integral family. For an integral $F(n_1, ..., n_9)$, we defined its sector as $[s_1, ..., s_9]$ with $s_i = \Theta(n_i - 1/2)$\footnote{The concept of sector is equivalent to the topology of Feynman diagrams, and thus we do not distinguish between sector and topology in this paper.}. These non-factorizable box-type two-loop Feynman diagrams can be categorized into $8$ topologies. The $4$ topologies shown in Figure \ref{fig1} belong to the integral family $\mathcal{F}$, which is identified by the set of propagators
      \begin{equation}
      \begin{aligned}
      D_1 &= l_1^2\,, \qquad &
      D_2 &= (l_1+p_1)^2-m_W^2\,, \qquad &
      D_3 &= (l_1-p_2)^2\,, \qquad \\
      D_4 &= l_2^2\,, \qquad &
      D_5 &= (l_2+p_1)^2\,, \qquad &
      D_6 &= (l_2-p_2)^2\,, \qquad \\
      D_7 &= (l_1-l_2)^2\,, \qquad &
      D_8 &= (l_1-p_2+p_3)^2 - m_{\ell}^2\,, \quad &
      D_9 &= (l_2-p_2+p_3)^2\,,
      \end{aligned}
      \end{equation}
      while the other $4$ topologies can be obtained from Figure \ref{fig1} by exchanging $p_1$ and $p_2$ and thus belong to the family $\mathcal{F}^{\ast} \equiv \mathcal{F}\big|_{p_1 \leftrightarrow p_2}$. The MIs for the non-factorizable mixed QCD-QED two-loop virtual corrections to $q\bar{q}^{\prime} \rightarrow \ell \nu$ will be computed analytically in Section \ref{sec:3}.
      \begin{figure}[H]
      \centering
      \includegraphics[width=0.9\textwidth]{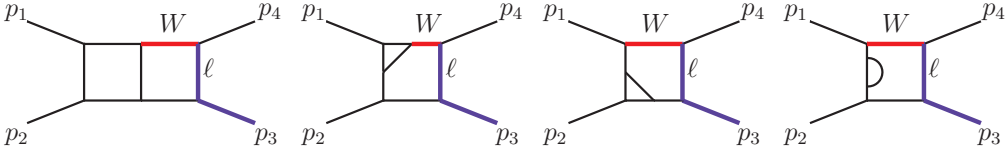}
      \caption{
      Four topologies of Feynman diagrams for non-factorizable mixed QCD-QED two-loop virtual corrections to $q\bar{q}^{\prime} \rightarrow \ell \nu$. All the diagrams in this paper are drawn by using \texttt{JaxoDraw} \cite{Binosi:2008ig}.}
      \label{fig1}
      \end{figure}
\item {\it Factorizable corrections}

      There are totally $24$ factorizable Feynman diagrams for the mixed QCD-QED two-loop virtual corrections to $q\bar{q}^{\prime} \rightarrow \ell \nu$. Four of them are reducible and can be evaluated analytically by using the techniques for one-loop integrals. The rest $20$ Feynman diagrams are irreducible and can be regarded as the mixed QCD-QED two-loop virtual corrections to $q\bar{q}^{\prime} \rightarrow W^{\ast}$. These two-loop three-point Feynman diagrams can be categorized into $8$ topologies which belong to $3$ different integral families. A detailed discussion on the MIs for the mixed QCD-QED two-loop virtual corrections to $q\bar{q}^{\prime} \rightarrow W^{\ast}$ is given in Appendix \ref{sec:appendix-A}.
\end{itemize}

\section{Non-factorizable mixed QCD-QED virtual corrections to $q\bar{q}^{\prime} \rightarrow \ell \nu$}
\label{sec:3}
As stated in Section \ref{sec:2}, the $8$ non-factorizable two-loop Feynman diagrams for the mixed QCD-QED virtual corrections to $q\bar{q}^{\prime} \rightarrow \ell \nu$ are categorized into $8$ topologies respectively, which belong to $2$ integral families. Due to the fact that the $4$ topologies/diagrams belonging to $\mathcal{F}^{\ast}$ can be obtained from the corresponding ones belonging to $\mathcal{F}$ by exchanging $p_1$ and $p_2$, only the $4$ topologies in Figure \ref{fig1} are considered in the following discussion.
\clearpage 
\par
The $4$ topologies in Figure \ref{fig1} belong to the integral family $\mathcal{F}$, corresponding to the following $4$ sectors of $\mathcal{F}$, respectively,
\begin{equation}
\begin{aligned}
\mbox{}[0, 1, 1, 1, 1, 1, 1, 1, 0]\,, \\
[1, 1, 1, 1, 1, 0, 1, 1, 0]\,, \\
[1, 1, 1, 1, 0, 1, 1, 1, 0]\,, \\
[1, 1, 1, 1, 0, 0, 1, 1, 0]\,.
\end{aligned}
\end{equation}
For a given sector $[s_1, s_2, ...]$, we define
\begin{equation}
[s_1, s_2, ...]_{\digamma}
~ =
\bigcup_{s_{i}^{\prime} \in \{0, 1\}\, \text{and}\, \leqslant s_i}[s_1^{\prime}, s_2^{\prime}, ...]\,.
\end{equation}
The scalar integrals induced by a Feynman diagram (corresponding a tensor integral) of this sector/topology via tensor reduction belong to $[s_1, s_2, ...]_{\digamma}$. Thus, the scalar integrals induced by the $4$ topologies in Figure \ref{fig1} certainly belong to the following integral set,
\begin{equation}
\mathcal{S}
=
[0, 1, 1, 1, 1, 1, 1, 1, 0]_{\digamma} \cup\,
[1, 1, 1, 1, 1, 0, 1, 1, 0]_{\digamma} \cup\,
[1, 1, 1, 1, 0, 1, 1, 1, 0]_{\digamma}\,.
\label{set-S}
\end{equation}
We compute the MIs of $\mathcal{S}$ in the unphysical region, namely $s < 0,~ -m_W^2 < t <0,~ m_{\ell}^2 > 0$ and $m_W^2 > 0$, where all the MIs are real-valued. Then their values in the physical region can be obtained by analytic continuation.

\subsection{Canonical differential equations}
\label{subsec:3-1}
\par
\begin{figure}[htbp]
\centering
\includegraphics[width=1.0\textwidth]{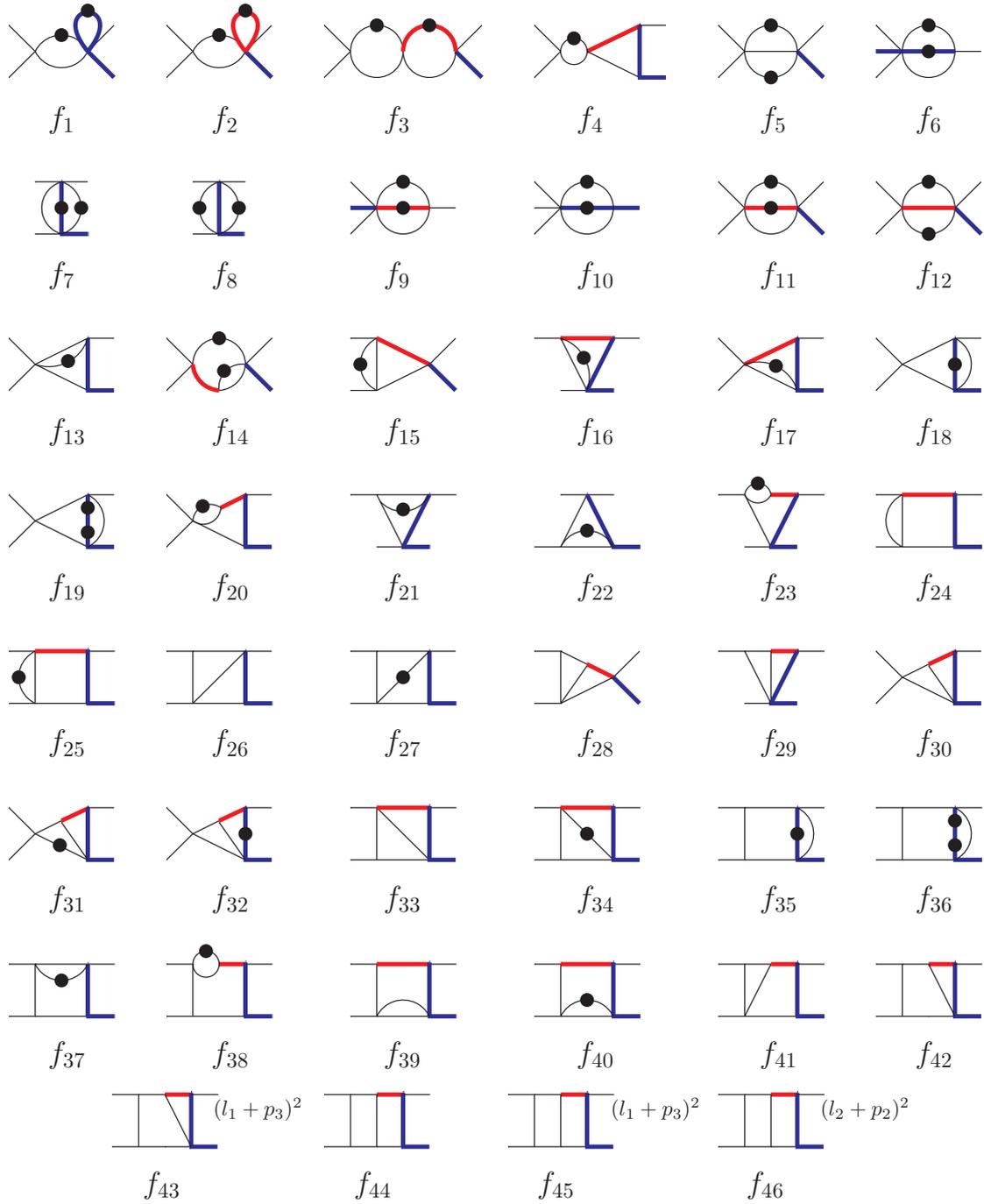}
\caption{A set of pre-canonical MIs. The thin and thick lines represent massless and massive external particles and propagators respectively (red for $W$ boson and blue for charged lepton). One dot means the index of that propagator is raised to $2$ and two dots go for $3$.}
\label{fig2}
\end{figure}

In this section, we elaborate on the construction of canonical differential equations for the MIs and present the solution of the canonical differential system. The scalar Feynman integrals of a given family are not independent: there are integration-by-parts (IBP) recurrence relations \cite{Chetyrkin:1981qh}. After performing IBP reduction procedure, a minimal number of irreducible integrals which are also called master integrals are obtained; any dimensionally regularized scalar integral in this family can be expressed as a linear combination of the MIs with coefficients being the rational functions of kinematic variables and spacetime dimension. This is quite important for using the differential equation method \cite{Kotikov:1990kg,Remiddi:1997ny} to compute scalar Feynman integrals. In this work, we adopt \texttt{Kira} \cite{Maierhoefer:2017hyi,Klappert:2020nbg} based on Laporta's algorithm \cite{Laporta:2000dsw} to perform IBP reduction, and then obtain $46$ MIs, $f_{1, ..., 46}$ of $\mathcal{S}$, which are depicted in Figure \ref{fig2}.

\par
The mass dimension of the scalar Feynman integral $F$ in Eq.(\ref{eq:FI}) is given by
\begin{equation}
[F] = 2 \left( d - \alpha \right)
\qquad
\text{with}
\qquad
\alpha = \sum_{i=1}^9 \alpha_i\,.
\end{equation}
Thus, $F$ can be nondimensionalized by multiplying a factor of $Q^{-[F]}$, i.e.,
\begin{equation}
F \quad\longrightarrow\quad Q^{-[F]} \times F\,,
\end{equation}
where $Q$ is a characteristic mass scale. In this work, we take $Q=m_W$ and introduce the following three dimensionless variables for convenience,
\begin{equation}
x=-\frac{s}{m_W^2}\,,
\qquad
y=-\frac{t}{m_W^2}\,,
\qquad
z=\frac{m_{\ell}^2}{m_W^2}\,.
\end{equation}
Then the nondimensionalized scalar Feynman integrals are functions of $x$, $y$ and $z$.

\par
With the help of \texttt{LiteRed} \cite{Lee:2012cn,Lee:2013mka}, we obtain the partial derivatives of the $46$ MIs with respect to $x$, $y$ and $z$. By means of IBP identities, those derivatives are expressed as the linear combinations of the $46$ MIs. This leads to the following set of partial differential equations for the vector of MIs $\textbf{F} \equiv (f_1, f_2, ..., f_{46})^{T}$,
\begin{equation}
\frac{\partial\textbf{F}}{\partial x} = \text{A}_x(x,y,z; \epsilon)\,\textbf{F}\,,
\qquad
\frac{\partial\textbf{F}}{\partial y} = \text{A}_y(x,y,z; \epsilon)\,\textbf{F}\,,
\qquad
\frac{\partial\textbf{F}}{\partial z} = \text{A}_z(x,y,z; \epsilon)\,\textbf{F}\,.
\end{equation}
It was suggested in Ref.\cite{Henn:2013pwa} that a good choice of MIs would simplify the calculation tremendously. By means of the following basis transformation $\mathbb{T}: \textbf{F} \longmapsto \textbf{G}$
\begin{align}
g_1    & = \epsilon^2\,f_1\,x\,,
       &   \qquad \qquad \quad
g_2    & = \epsilon^2\,f_2\,x\,,
           \nonumber \\
g_3    & = \epsilon^2\,f_3\,x^2\,,
       &
g_4    & = \epsilon^3\,f_4\,x\,(x+z)\,,
           \nonumber \\
g_5    & = \epsilon^2\,f_5\,x\,,
       &
g_6    & = \epsilon\,(1-\epsilon)\,f_6\,z\,,
           \nonumber \\
g_7    & = \epsilon^2\,f_7\,y\,,
       &
g_8    & = \epsilon^2\,f_8\,(y+z) + 2\,\epsilon^2\,f_7\,z\,,
           \nonumber \\
g_9    & = \epsilon\,(1-\epsilon)\,f_9\,,
       &
g_{10} & = \epsilon^2\,f_{10}\,z\,,
           \nonumber \\
g_{11} & = \epsilon^2\,f_{11}\,x\,,
       &
g_{12} & = \epsilon^2\,f_{12}\,(x+1) + 2\,\epsilon^2\,f_{11}\,,
           \nonumber \\
g_{13} & = \epsilon^3\,f_{13}\,(x+z)\,,
       &
       &
           \nonumber \\
g_{14} & = \rlap{$\displaystyle
           [\,
           \epsilon^2\,f_{14}\,x\,(x+1)
	       + (1/2)\,\epsilon\,(1-\epsilon)\,f_9\,x
	       + (3/2)\,\epsilon^2\,f_5\,x^2
           \,]/(1-x)\,,
	       $} \nonumber \\
g_{15} & = \epsilon^3\,f_{15}\,x\,,
       &
g_{16} & = \epsilon^3\,f_{16}\,y\,,
           \nonumber \\
g_{17} & = \epsilon^3\,f_{17}\,(x+z)\,,
       &
g_{18} & = \epsilon^3\,f_{18}\,(x+z)\,,
           \nonumber \\
g_{19} & = \epsilon^2\,f_{19}\,z\,(x+z)\,,
       &
g_{20} & = \epsilon^3\,f_{20}\,(x+z)\,,
           \nonumber \\
g_{21} & = \epsilon^3\,f_{21}\,y\,,
       &
g_{22} & = \epsilon^3\,f_{22}\,(y+z)\,,
\label{eq:UTbasis}
           \\
g_{23} & = \epsilon^3\,f_{23}\,y\,,
       &
g_{24} & = \epsilon^3\,(1-2\epsilon)\,f_{24}\,(x+z)\,,
           \nonumber \\
g_{25} & = \epsilon^3\,f_{25}\,(x+1)\,(y+z)\,,
       &
g_{26} & = \epsilon^4\,f_{26}\,(x+y)\,,
           \nonumber \\
g_{27} & = \epsilon^3\,f_{27}\,x\,(y+z)\,,
       &
g_{28} & = \epsilon^4\,f_{28}\,x\,,
           \nonumber \\
g_{29} & = \epsilon^4\,f_{29}\,y\,,
       &
g_{30} & = \epsilon^4\,f_{30}\,(x+z)\,,
           \nonumber \\
g_{31} & = \epsilon^3\,f_{31}\,x\,(x+z)\,,
       &
g_{32} & = \epsilon^3\,f_{32}\,(1-z)\,(x+z)\,,
           \nonumber \\
g_{33} & = \epsilon^4\,f_{33}\,(x+y+z)\,,
       &
g_{34} & = \epsilon^3\,f_{34}\,(x+1)\,(y+z)\,,
           \nonumber \\
g_{35} & = \epsilon^3\,f_{35}\,x\,y\,,
       &
g_{36} & = \epsilon^2\,[\, f_{36}\,(y+z) + \epsilon\,f_{35} \,]\,x\,z\,,
           \nonumber \\
g_{37} & = \epsilon^3\,f_{37}\,x\,(y+z)\,,
       &
g_{38} & = \epsilon^3\,[\, f_{38}\,(x+1) + f_{37} \,]\,(y+z)\,,
           \nonumber \\
g_{39} & = \epsilon^3\,(1-2 \epsilon)\,f_{39}\,y\,,
       &
g_{40} & = \epsilon^3\,f_{40}\,(x+1)\,(y+z)\,,
           \nonumber \\
g_{41} & = \epsilon^4\,f_{41}\sqrt{\lambda}\,,
       &
g_{42} & = \epsilon^4\,f_{42}\,x\,y\,,
           \nonumber \\
g_{43} & = \epsilon^4\,f_{43}\,x - \epsilon^4\,f_{33}\,(2 x + y + z)\,,
       &
g_{44} & = \epsilon^4\,f_{44}\,x\,(x+1)\,(y+z)\,,
           \nonumber \\
g_{45} & = \epsilon^4\,(f_{45}\,x - f_{44}\,x\,z + f_{41}\,y)\,(x+1)\,,
       &
g_{46} & = \epsilon^4\,f_{46}\,x\,(x+z)\,,
           \nonumber
\end{align}
we obtain a canonical set of MIs $\textbf{G} = (g_1, g_2, ..., g_{46})^{T}$. Such a canonical basis $\textbf{G}$ obeys the following system of differential equations,
\begin{equation}
\frac{\partial\textbf{G}}{\partial x} = \epsilon\,\mathbb{A}_x(x,y,z)\,\textbf{G}\,,
\qquad
\frac{\partial\textbf{G}}{\partial y} = \epsilon\,\mathbb{A}_y(x,y,z)\,\textbf{G}\,,
\qquad
\frac{\partial\textbf{G}}{\partial z} = \epsilon\,\mathbb{A}_z(x,y,z)\,\textbf{G}\,,
\label{eq:epform}
\end{equation}
where the dependence on the dimensional regularization parameter $\epsilon$ is completely factorized from the kinematics. Since the canonical differential system in Eq.(\ref{eq:epform}) is invariant under the scaling transformation $\textbf{G} \longmapsto \epsilon^{a} \textbf{G}$, the canonical basis $\textbf{G}$ defined in Eq.(\ref{eq:UTbasis}) has been required to be analytic and nonzero at $\epsilon = 0$. Finally, it is worth mentioning that all the entries of the basis transformation matrix $\mathbb{T}$ are rational functions of $x$, $y$ and $z$, except
\begin{equation}
\label{eq-root}
\mathbb{T}_{41,41} = \epsilon^4 \sqrt{\lambda}
\qquad
\text{with}
\qquad
\lambda= \big[ x\,(y+z)+ x + y \big]^2 - 4\,x^2\,z\,.
\end{equation}

\subsection{Integrations}
\label{subsec:3-2}
The partial differential equations in Eq.(\ref{eq:epform}) can be combined into a $d\log$-form total differential equation,
\begin{equation}
d \textbf{G} = \epsilon\,d \mathbb{A}\,\textbf{G}\,,
\qquad
d \mathbb{A} = \sum_{i=1}^{15} \mathbb{C}_i\, d \log \eta_i\,,
\end{equation}
where $\mathbb{C}_i~ (i = 1, ..., 15)$ are constant matrices with rational-number entries, and the arguments of the $d\log$'s,
\begin{align}
   \eta_1 &= x\,, &
   \eta_6 &= x+y\,, &
   \eta_{11} &= y+z+x\,z\,,
\nonumber \\
   \eta_2 &= y\,, &
   \eta_7 &= x+z\,, &
   \eta_{12} &= \sqrt{\lambda}\,,
\nonumber \\
   \eta_3 &= z\,, &
   \eta_8 &= y+z\,, &
   \eta_{13} &= x\,(y+z)+x+y+\sqrt{\lambda}\,,
\label{eq:eta-15}
\\
   \eta_4 &= 1+x\,, &
   \eta_9 &= x+y+z\,, &
   \eta_{14} &= x\,(y+z)-x+y+\sqrt{\lambda}\,,
\nonumber \\
   \eta_5 &= 1-z\,, &
   \eta_{10} &= 1-y-z\,, &
   \eta_{15}& = x\,(y+z)-x-y+\sqrt{\lambda}\,,
\nonumber
\end{align}
are called letters and together constitute the alphabet of our problem. Considering the mass hierarchy between $W$ boson and charged lepton ($m_{\ell} \ll m_W$), we may expand the differential system for small $z$. After the expansion for small $z$, the coefficient matrix $d \mathbb{A}$ can be expressed as
\begin{equation}
d \mathbb{A}
=
\sum_{i=1}^{7} \widetilde{\mathbb{C}}_i\, d \log \alpha_i
+
d \left[\,
\sum_{n=1}^{+\infty}\, \mathbb{A}_n(x, y)\, z^n
\right]\,,
\label{eq:expansion-dA}
\end{equation}
where
\begin{equation}
\begin{aligned}
\alpha_1 & = x\,,        &\qquad \alpha_2 & = 1+x\,, &\qquad \alpha_3 & = x+y\,, \\
\alpha_4 & = x+y+x\,y\,, &\qquad \alpha_5 & = y\,,   &\qquad \alpha_6 & = 1-y\,, \\
\alpha_7 & = z\,,
\end{aligned}
\label{eq:alphabet}
\end{equation}
\begin{equation}
\begin{aligned}
\widetilde{\mathbb{C}}_1 & = \mathbb{C}_1 + \mathbb{C}_7 + \mathbb{C}_{15}\,,
& \qquad
\widetilde{\mathbb{C}}_2 & = \mathbb{C}_4 + \mathbb{C}_{14}\,,
& \qquad
\widetilde{\mathbb{C}}_3 & = \mathbb{C}_6 + \mathbb{C}_9\,,
\\
\widetilde{\mathbb{C}}_4 & = \mathbb{C}_{12} + \mathbb{C}_{13}\,,
& \qquad
\widetilde{\mathbb{C}}_5 & =
\rlap{$\mathbb{C}_2 + \mathbb{C}_8 + \mathbb{C}_{11} + \mathbb{C}_{14} + \mathbb{C}_{15}\,,$}
\\
\widetilde{\mathbb{C}}_6 & = \mathbb{C}_{10}\,,
& \qquad
\widetilde{\mathbb{C}}_7 & = \mathbb{C}_3\,,
\end{aligned}
\end{equation}
and $\mathbb{A}_n(x, y)~ (n \in \mathbb{N}^+)$ are completely comprised of rational-function entries\footnote{All the poles of $\mathbb{A}_n(x, y)~ (n \in \mathbb{N}^+)$ are zeros of $\alpha_i~ (i = 1, ..., 6)$.}. Therefore, the matrices $\mathbb{A}_{x,y}(x, y, z)$ and $\mathbb{A}_z(x, y, z)$ in Eq.(\ref{eq:epform}) are analytic and singular at $z = 0$, respectively, and can be expanded as
\begin{equation}
\left\lbrace
~
\begin{aligned}
&
\begin{aligned}
\mathbb{A}_{x}(x, y, z) &= \sum_{n=0}^{+\infty}\, \mathbb{A}_{x,n}(x, y)\, z^n
\\
\mathbb{A}_{y}(x, y, z) &= \sum_{n=0}^{+\infty}\, \mathbb{A}_{y,n}(x, y)\, z^n
\end{aligned}
&\quad&
\text{(Taylor expansion)}
\\
&
\mathbb{A}_{z}(x, y, z) = \frac{\mathbb{A}_{z,-1}}{z} + \sum_{n=0}^{+\infty}\, \mathbb{A}_{z,n}(x, y)\, z^n
&\quad&
\text{(Laurent expansion)}
\end{aligned}
\right.
\label{eq:expansion-Axyz}
\end{equation}
In the lowest-order approximation,
\begin{equation}
d \mathbb{A}
\simeq
\sum_{i=1}^{7} \widetilde{\mathbb{C}}_i\, d \log \alpha_i\,,
\end{equation}
the $7$ letters $\alpha_i~ (i = 1, ..., 7)$ constitute the alphabet of the differential system. As is well known, the leading terms in the $z$-expansion in Eq.(\ref{eq:expansion-Axyz}) ($\mathbb{A}_{x,0}$, $\mathbb{A}_{y,0}$ and $\mathbb{A}_{z,-1}$) are responsible for the logarithmic lepton-mass singularities of the canonical MIs $\textbf{G}$. Thus, the logarithmically divergent contributions of the form $\log^n z$ can be fully captured by adopting the lowest-order approximation. In this work, we adopt the second-order approximation, i.e.,
\begin{equation}
d \mathbb{A}
\simeq
\sum_{i=1}^{7} \widetilde{\mathbb{C}}_i\, d \log \alpha_i
+
d \big[
\mathbb{A}_1(x, y)\,z
\big]\,,
\label{eq:2nd-order}
\end{equation}
in order to get the analytic expression for $\textbf{G}$ up to $\mathcal{O}(z\log^n z)$. In the unphysical region,
\begin{equation}
x > 0 ~ \wedge ~ 0 < y < 1 ~ \wedge ~ z > 0\,,
\end{equation}
$\alpha_i~ (i = 1, ..., 7)$ are real and positive. To evaluate the MIs in the physical region, the same analytic continuation procedure as in Ref.\cite{Bonciani:2016ypc} should be applied.

\par
Since $\textbf{G}$ is finite in the limit $\epsilon \rightarrow 0$, it can be expanded in a Taylor series around four dimensions,
\begin{equation}
	\textbf{G} = \sum_{n=0}^{\infty} \textbf{G}^{(n)}\,\epsilon^n\,.
\end{equation}
From Eq.(\ref{eq:epform}), one can establish the differential equations connecting $\textbf{G}^{(n)}$ and $\textbf{G}^{(n-1)}$. These first-order differential equations can be integrated in an iterative manner, namely order by order in $\epsilon$. Then the canonical MIs at $\mathcal{O}(\epsilon^n)$ can be expressed as
\begin{equation}
	\textbf{G}^{(n)} = \sum_{m=0}^n \mathbb{M}^{(n-m)}\,\textbf{C}^{(m)}\,,
\end{equation}
where $\textbf{C}^{(n)}$ is the integration constant vector at $\mathcal{O}(\epsilon^n)$ and the matrices $\mathbb{M}^{(n)}~ (n \in \mathbb{N})$ are defined recursively by
\begin{equation}
\mathbb{M}^{(n)}=\mathbb{M}_x^{(n)} + \mathbb{M}_y^{(n)} + \mathbb{M}_z^{(n)}\,,
\qquad
\left\lbrace
\begin{aligned}
~\mathbb{M}_x^{(n)}&=\int\Ax\,\mathbb{M}^{(n-1)}\,dx \\
~\mathbb{M}_y^{(n)}&=\int\left(\Ay\,\mathbb{M}^{(n-1)} - \partial_y\,\mathbb{M}_x^{(n)}\right)\,dy \\
~\mathbb{M}_z^{(n)}&=\int\left(\Az\,\mathbb{M}^{(n-1)} - \partial_z\,\mathbb{M}_x^{(n)} - \partial_z\,\mathbb{M}_y^{(n)} \right)\,dz
\end{aligned}
\right.
\label{eq:Mmat}
\end{equation}
with $\mathbb{M}^{(0)}=\mathbf{1}$. Actually, it is sufficient to expand $\textbf{G}$ up to the order of $\epsilon^4$ for the mixed QCD-QED two-loop virtual corrections to charged-current DY processes. Since the entries of $\mathbb{A}_{x,\,y,\,z}$ are all rational functions of $x$, $y$ and $z$ after the expansion for small $z$, $\mathbb{M}^{(n)}$ can be expressed in terms of rational functions and Goncharov polylogarithms (GPLs) \cite{Goncharov:2001iea, Goncharov:1998kja}. A GPL with weights $w_i~ (i = 1, ..., n)$ and argument $t$ is defined recursively by
\begin{equation}
G(w_n, ..., w_1; t) = \int_0^t \frac{1}{\tau - w_n}\,G(w_{n-1}, ..., w_1; \tau)\,d\tau
\label{eq:GPLdef}
\end{equation}
with $G(\,; t) = 1$ and
\begin{equation}
G(\vec{0}_n; t) = \frac{\log^n t}{n!}\,.
\end{equation}
For the MIs $\textbf{G}$ involved in the calculation of the non-factorizable mixed QCD-QED two-loop virtual corrections to $qq^{\prime} \rightarrow \ell \nu$, the weights of the GPLs with arguments $x$, $y$ and $z$ are drawn from the following three sets, respectively,
\begin{equation}
\mathcal{W}_x = \Big\{ 0,\, -1,\, -y,\, -\frac{y}{1+y} \Big\}\,,
\qquad
\mathcal{W}_y = \{ 0,\, 1 \}\,,
\qquad
\mathcal{W}_z = \{ 0 \}\,.
\end{equation}
In the recursive definition for $\mathbb{M}^{(n)}$ in Eq.(\ref{eq:Mmat}), one encounters the integrals with integrands of the form $r(t)\,G(\vec{w}; t)$, where $r(t)$ and $G(\vec{w}; t)$ represent rational functions and GPLs, respectively. Those integrals can be calculated recursively by using the following IBP recurrence relation,
\begin{equation}
\int_0^t r(\tau)\,G(\vec{w}_n; \tau)\,d\tau
=
\Big[
R(\tau)\,G(\vec{w}_n; \tau)
\Big]_0^t
-
\int_0^t R(\tau)\,\frac{1}{\tau - w_n}\,G(\vec{w}_{n-1}; \tau)\,d\tau\,,
\end{equation}
where $\vec{w}_n = (w_n, ..., w_1)$, $\vec{w}_{n-1} = (w_{n-1}, ..., w_1)$, and $R(t)$ is an antiderivative of $r(t)$. In this paper we employ the \texttt{Mathematica} package \texttt{PolyLogTools} \cite{Maitre:2005uu,Maitre:2007kp,Duhr:2019tlz} and \texttt{C++} library \texttt{GiNaC} \cite{Bauer:2000cp} for both symbolic computation and numerical evaluation \cite{Vollinga:2004sn} of GPLs.

\subsection{Boundary conditions}
\label{subsec:3-3}
\par
To obtain the exact solution of the canonical differential system in Eq.(\ref{eq:epform}), the integration constants (i.e., the boundary terms of the MIs $g_{1,...,46}$) have to be specified. For some of the MIs, we take their known analytic solutions from the literature or evaluate them by means of some much simpler differential systems. For all other MIs, the boundary terms are fixed by imposing a set of regularity conditions, i.e., demanding regularity of those MIs or their linear combinations in certain kinematic limits. It was found that the weight structure of integration constants in Ref.\cite{Hasan:2020vwn} is quite simple. The integration constants at $\mathcal{O}(\epsilon^{n})~ (n = 0, 1, 2, 3, 4)$ are proportional to $1$, $0$, $\zeta(2)$, $\zeta(3)$ and $\zeta(4)$, respectively. This property still holds in our case. For our problem, the conditions imposed to the MIs for the determination of their boundary constants are listed below.
\begin{itemize}
\item $g_{1,2,3,5,...,12,14,15,22,28}$ are taken from Refs.\cite{Bonciani:2016ypc,Hasan:2020vwn} as independent input.
\item $g_{13,18}$ are determined by matching against the MIs with full lepton mass dependence, which are computed in Appendix \ref{sec:appendix-B}.
\item The boundary constants of the rest 29 MIs are fixed by the following linearly independent regularity conditions\footnote{The asymptotic behaviour of $\textbf{G}$ in the limit $\alpha_{i} \rightarrow 0~ (i = 1, ..., 7)$ is provided in Appendix \ref{sec:appendix-C}.}:
    \begin{itemize}
    \item regularity at $x \rightarrow 0$:\, $g_{4,17,24,25,30,31,34,40,46}$
    \item regularity at $x \rightarrow -y$:\, $g_{26,27,33,35,36,37,38,40,41,42,43,44}$
    \item regularity at $x \rightarrow -1$:\, $g_{32}$
    \item regularity at $x \rightarrow -y/(1+y)$:\, $g_{41,44}$
    \item regularity at $y \rightarrow 1$:\, $g_{16,23}$
    \item regularity at $z \rightarrow 0$ and matching against their massless ($m_{\ell} = 0$) counterparts: 
        \begin{equation}
        \left\lbrace\,
        \begin{aligned}
        h_a &= \frac{g_1}{2}+g_4+g_{25}-g_{37}-g_{38}+g_{45} \\
        h_b &= g_{30} \\
        h_c &= \frac{g_6}{4}+g_{21}
        \end{aligned}
        \right.
        \label{eq:habc}
        \end{equation}
    \end{itemize}
\end{itemize}
The boundary conditions for $h_{a,b,c}$ are determined from the $m_{\ell} \rightarrow 0$ behaviour of these integrals, where all the internal lepton propagators become massless. Around the $z = 0$ singularity, the differential equation for the $z$-evolution reduces to
\begin{equation}
\frac{\partial\textbf{G}}{\partial z} \simeq \epsilon\,\frac{\mathbb{A}_{z,-1}}{z}\,\textbf{G}\,.
\end{equation}
To figure out the $m_{\ell} \rightarrow 0$ behaviour of the $46$ MIs, we perform a Jordan decomposition of the pole matrix $\mathbb{A}_{z,-1}$ \cite{Mastrolia:2017pfy}, i.e.,
\begin{equation}
	\mathbb{A}_{z,-1} = \mathbb{S}^{-1} \mathbb{J}\,\mathbb{S}\,,
\end{equation}
where $\mathbb{J}$ is the Jordan canonical form of $\mathbb{A}_{z,-1}$,
\begin{equation}
\mathbb{J} =
\left(

\right)\,,
\end{equation}
and the similarity transformation $\mathbb{S}$ reads
\begin{equation}
\hspace{-1.5em}
\scalemath{0.67}{
\left(\,
%
~\right)
}
\label{eq:S-matrix}
\end{equation}
Then the $z$-evolution around the $z=0$ singularity can be described in Jordan form,
\begin{equation}
\frac{\partial\textbf{H}}{\partial z} \simeq \epsilon\,\frac{\mathbb{J}}{z}\,\textbf{H}\,,
\label{eq:z-evolution-J}
\end{equation}
where $\textbf{H} = (h_1, h_2, ..., h_{46})^{T}$ is the Jordan canonical basis defined by $\textbf{H} = \mathbb{S} \, \textbf{G}$.
Considering the Jordan block structure of $\mathbb{J}$, the singular behaviour of $\textbf{H}$ in the limit $z \rightarrow 0$ can be simply factored out, and the solution of Eq.(\ref{eq:z-evolution-J}) can be expressed as
\begin{equation}
\textbf{H} = \mathbb{R}^{-1} \textbf{H}_{0}\,,
\qquad\quad
\mathbb{R}
=
z^{2\epsilon}\, \mathbbm{1}_{3 \times 3}
\oplus
z^{2\epsilon}
\left(
\begin{matrix}
~1~ & ~-\epsilon \log z \\
~0~ & ~~1
\end{matrix}
\right)
\oplus
z^{\epsilon}\, \mathbbm{1}_{9 \times 9}
\oplus
\mathbbm{1}_{32 \times 32}\,,
\end{equation}
where $\textbf{H}_{0} = (h_{1,0}, h_{2,0}, ..., h_{46,0})^{T}$ is analytic at $z = 0$. For convenience, we define
\begin{equation}
\widetilde{\textbf{H}} = (h_1, h_2, ..., h_{14})^{T}\,,
\qquad
\widehat{\textbf{H}} = (h_{15}, h_{16}, ..., h_{46})^{T}\,,
\end{equation}
and
\begin{equation}
\widetilde{\mathbb{R}}
=
z^{2\epsilon}\, \mathbbm{1}_{3 \times 3}
\oplus
z^{2\epsilon}
\left(
\begin{matrix}
~1~ & ~-\epsilon \log z \\
~0~ & ~~1
\end{matrix}
\right)
\oplus
z^{\epsilon}\, \mathbbm{1}_{9 \times 9}\,.
\end{equation}
The regularity of $\mathbb{R} \textbf{H}$ at $z \rightarrow 0$ can be applied to the following two aspects.
\begin{enumerate}
\item {\it Boundary conditions:} \\
      \begin{equation}
      \lim_{z \rightarrow 0} \widehat{\textbf{H}} = \mathbb{K}\, \textbf{I}\,.
      \label{eq:Hhat}
      \end{equation}
      $\widehat{\textbf{H}}$ is analytic at $z=0$, the l.h.s of Eq.(\ref{eq:Hhat}) can be computed by taking the limit $z \rightarrow 0$ directly at the integrand level and thus expressed in terms of the canonical basis $\textbf{I} = (\text{I}_1, \text{I}_2, ..., \text{I}_{31})^T$ of the two-loop scalar integrals involved in the mixed QCD-QED virtual corrections to the charged-current Drell-Yan production of a massless charged lepton \cite{Bonciani:2016ypc}.
      \par
      In Eq.(\ref{eq:habc}) we introduced three two-loop scalar integrals $h_a$, $h_b$ and $h_c$ for fixing the boundary terms of $\textbf{G}$. Considering the analytic expression of $\mathbb{S}$ in Eq.(\ref{eq:S-matrix}), $h_{a, b, c}$ can be written as linear combinations of $\widehat{\textbf{H}}$,
      \begin{equation}
      \begin{aligned}
      h_a &= h_{16}-h_{23}-h_{24}+h_{35}+h_{45}\,, \\
      h_b &= h_{30}\,, \\
      h_c &= h_{39}\,,
      \end{aligned}
      \end{equation}
      and thus are finite in the limit $z \rightarrow 0$. After performing IBP reduction, we obtain
      \begin{equation}
      \begin{aligned}
      \lim_{z \rightarrow 0}\, h_a &= -\,\text{I}_{11} - \text{I}_{17} + \text{I}_{23} - \text{I}_{26} + \text{I}_{29} + \text{I}_{31}\,, \\
      \lim_{z \rightarrow 0}\, h_b &= \text{I}_{13}\,, \\
      \lim_{z \rightarrow 0}\, h_c &= \text{I}_8\,.
      \end{aligned}
      \label{eq:h-I}
      \end{equation}
      The matching of $h_{a, b, c}$ to their massless counterparts, i.e., Eq.(\ref{eq:h-I}), provides three boundary conditions for $\textbf{G}$.
\item {\it Consistency checks:}
      \begin{equation}
      \widetilde{\mathbb{R}}\, \widetilde{\textbf{H}}\text{: finite in the limit $z \rightarrow 0$}
      \end{equation}
      Even though $z = 0$ is a singularity of $\widetilde{\textbf{H}}$, $\widetilde{\mathbb{R}} \widetilde{\textbf{H}}$ is finite when $z \rightarrow 0$.
      Thus, the finiteness of the following $14$ combinations of $\widetilde{\textbf{H}}$ in the zero lepton mass limit provides a consistency check for the analytic solution of the canonical differential system (\ref{eq:epform}),
      \begin{equation}
      \begin{aligned}
      & z^{2\epsilon}\, h_{1,2,3,5}\,, \\
      & z^{\epsilon}\, h_{6,7,8,9,10,11,12,13,14}\,, \\
      & z^{2\epsilon}\left(h_4 - \epsilon \log z\, h_5\right)\,.
      \end{aligned}
      \label{eq:H14}
      \end{equation}
\end{enumerate}

\subsection{Solution and checks}
\label{subsec:3-4}
\par
We obtain the solution of the canonical differential system (\ref{eq:epform}) in terms of GPLs of arguments $x$, $y$ and $z$ with weights $\{0,\, -1,\, -y,\, -y/(1+y)\}$, $\{0,\, 1\}$ and $\{0\}$, respectively. The analytic expressions of all the $46$ canonical MIs up to $\mathcal{O}(\epsilon^4)$ in the second-order approximation (i.e., Eq.(\ref{eq:2nd-order})) are provided in electronic form via an ancillary file on the $\texttt{arXiv}$. Since the weight of GPLs with argument $z$ is zero, the lepton-mass singularities of $\textbf{G}$ must be the logarithms of the form $\log^n (m_{\ell}^2/m_W^2)$ which can be completely determined by the leading terms in the $z$-expansion of $\mathbb{A}_{x, y, z}$. In Appendix \ref{sec:app3}, we showcase the explicit expressions of $g_{i}~ (i = 1, ..., 46)$ up to $\mathcal{O}(\epsilon^3)$ in the lowest-order approximation, i.e., setting $\{ \mathbb{A}_x,\, \mathbb{A}_y,\, \mathbb{A}_z \} = \{ \mathbb{A}_{x,0},\, \mathbb{A}_{y,0},\, \mathbb{A}_{z,-1}/z \}$. In order to verify our analytic solution we performed a number of checks:
\begin{itemize}
\item We checked our results in the Euclidean region $\{\,(x, y, z)\, |\, x > 0,\, 0 < y < 1,\, z > 0 \,\}$ against the numerical values obtained with \texttt{pySecDec} \cite{Borowka:2017idc, Borowka:2018goh} based on the sector decomposition algorithm. A comparison between the numerical results obtained from our analytic expressions and \texttt{pySecDec} for some representative MIs at $s = -5$, $t = -2$, $m_{\ell}^2 = 10^{-3}$ and $m_W^2 = 4~ \text{GeV}^2$ is presented in Table \ref{tab1}. It shows that all these numerical results are in good agreement with each other within the calculation errors.
\item The canonical basis $\boldsymbol{\Gamma} = (\gamma_1, \gamma_2, ..., \gamma_7)^{T}$ for the integral set $\mathcal{S}_{\text{aux}}$ defined by Eq.(\ref{set-Saux}) was calculated analytically in Appendix \ref{sec:appendix-B}. As a corollary of Eq.(\ref{eq:G-Gamma}), the MIs $g_{1},\, g_{5},\, g_{6},\, g_{10},\, g_{13},\, g_{18},\, g_{19}$, as functions of $x$ and $z$, must satisfy the following identities:
    \begin{equation}
    g_{1,5,6,10,13,18,19}(x=1, z) = \gamma_{1,2,4,3,7,5,6}(z)\,.
    \end{equation}
    These identities were verified at the analytical level in the lowest-order approximation up to $\mathcal{O}(\epsilon^4)$\footnote{The analytic expressions for $\gamma_{i}~ (i = 1, ..., 7)$ in the lowest-order approximation can be obtained from Eq.(\ref{eq:MIs-Gamma}) by setting the GPLs with non-zero weight to zero.}.
\item The $m_{\ell} \rightarrow 0$ behaviour of the Jordan canonical basis $\textbf{H} = \mathbb{S} \, \textbf{G}$ was checked.
    \begin{itemize}
    \item As stated in Section \ref{subsec:3-3}, $h_i~ (i = 15, ..., 46)$ are finite in the limit $z \rightarrow 0$, and their massless counterparts can be expressed as the linear combinations of $\text{I}_1$, $\text{I}_2$, ..., and $\text{I}_{31}$. The correctness of Eq.(\ref{eq:Hhat}) was confirmed by employing IBP recurrence relations and, consequently, the explicit expression of the $32 \times 31$ matrix $\mathbb{K}$ was obtained.
    \item $h_i~ (i = 1, ..., 14)$ are divergent as $m_{\ell} \rightarrow 0$, but the $14$ combinations of them defined by Eq.(\ref{eq:H14}) are finite in the zero lepton mass limit. The asymptotic behaviour of these combinations in the limit $z \rightarrow 0$ was checked both analytically and numerically, which clearly demonstrate the regularity of these combinations at $z = 0$.
    \end{itemize}
\end{itemize}
\begin{equation}
\mathbb{K}
\,=\,
\scalemath{0.67}{
\left(
\begin{matrix}
  \minus \frac{15}{8}       & \pminus \frac{1}{8}       & \pminus \frac{7}{16}      & \pminus \frac{1}{2}
& \pminus 0   & \pminus \frac{1}{16}      & \minus \frac{3}{2}        & \pminus 0   & \pminus 3   & \minus 1
& \pminus 0   & \pminus \frac{1}{4}       & \pminus 0   & \minus \frac{1}{2}        & \pminus 0
& \pminus \frac{1}{2}       & \pminus 0   & \minus \frac{1}{2}        & \pminus 0   & \pminus \frac{3}{4}
& \minus \frac{1}{8}        & \pminus 0   & \pminus 0   & \pminus 0   & \pminus 0   & \pminus 0
& \pminus \frac{1}{2}       & \minus \frac{1}{2}        & \pminus 0   & \pminus 0   & \pminus 1
\\
  \pminus 0   & \pminus 0   & \pminus 0   & \pminus 0   & \pminus 0   & \pminus 0   & \pminus 0   & \pminus 0
& \pminus 0   & \pminus 0   & \pminus 0   & \pminus 0   & \pminus 0   & \pminus 0   & \pminus 0   & \pminus 0
& \pminus 0   & \pminus 0   & \pminus 0   & \pminus 0   & \pminus 0   & \pminus 0   & \pminus 0   & \pminus 0
& \pminus 0   & \pminus 0   & \pminus 0   & \pminus 0   & \pminus 1   & \pminus 0   & \pminus 1
\\
  \pminus 0   & \pminus 0   & \pminus 0   & \pminus 0   & \pminus 0   & \pminus 0   & \pminus 0   & \pminus 0
& \pminus 0   & \pminus 0   & \pminus 0   & \pminus 0   & \pminus 0   & \pminus 0   & \pminus 0   & \pminus 0
& \pminus 0   & \pminus 0   & \pminus 0   & \pminus 0   & \pminus 0   & \pminus 0   & \pminus 0   & \pminus 0
& \pminus 0   & \pminus 0   & \pminus 0   & \pminus 0   & \pminus 0   & \pminus 1   & \pminus 0
\\
  \pminus \frac{7}{8}       & \pminus \frac{7}{8}       & \minus \frac{7}{16}       & \minus \frac{1}{2}
& \pminus 0   & \minus \frac{1}{16}       & \pminus \frac{1}{2}       & \pminus 0   & \pminus 0   & \pminus 0
& \pminus 0   & \minus \frac{1}{4}        & \pminus 0   & \minus \frac{1}{2}        & \pminus 0
& \minus \frac{1}{6}        & \pminus 0   & \pminus \frac{1}{2}       & \pminus 0   & \minus \frac{7}{4}
& \pminus \frac{1}{8}       & \pminus 0   & \pminus 0   & \pminus 0   & \pminus 0   & \pminus 0
& \minus \frac{1}{2}        & \pminus \frac{1}{2}       & \pminus 0   & \pminus 0   & \pminus 0
\\
  \pminus 0   & \pminus 0   & \pminus 0   & \pminus 0   & \pminus 0   & \pminus 0   & \pminus 0   & \pminus 0
& \pminus 0   & \pminus 0   & \pminus 0   & \pminus 0   & \pminus 0   & \pminus 0   & \pminus 0   & \pminus 0
& \pminus 0   & \pminus 0   & \pminus 0   & \pminus 0   & \pminus 0   & \pminus 0   & \pminus 0   & \pminus 0
& \pminus 0   & \pminus 0   & \pminus 1   & \pminus 0   & \pminus 0   & \pminus 0   & \pminus 0
\\
  \pminus 0   & \pminus 0   & \pminus 0   & \pminus 0   & \pminus 0   & \pminus 0   & \pminus 0   & \pminus 0
& \pminus 0   & \pminus 0   & \pminus 0   & \pminus 0   & \pminus 0   & \pminus 0   & \pminus 0   & \pminus 0
& \pminus 0   & \pminus 0   & \pminus 0   & \pminus 0   & \pminus 0   & \pminus 0   & \pminus 0   & \pminus 0
& \pminus 0   & \pminus 0   & \pminus 0   & \pminus 0   & \pminus 1   & \pminus 0   & \pminus 0
\\
  \pminus 0   & \pminus 0   & \pminus 0   & \pminus 0   & \pminus 0   & \pminus 0   & \pminus 0   & \pminus 0
& \pminus 0   & \pminus 0   & \pminus 0   & \pminus 0   & \pminus 0   & \pminus 0   & \pminus 0   & \pminus 0
& \pminus 0   & \pminus 0   & \pminus 0   & \pminus 0   & \pminus 0   & \pminus 0   & \pminus 0   & \pminus 0
& \pminus 1   & \pminus 0   & \pminus 0   & \pminus 0   & \pminus 0   & \pminus 0   & \pminus 0
\\
  \pminus 0   & \pminus 0   & \pminus 0   & \pminus 0   & \pminus 0   & \pminus 0   & \pminus 0   & \pminus 0
& \pminus 0   & \pminus 0   & \pminus 0   & \pminus 0   & \pminus 0   & \pminus 0   & \pminus 0   & \pminus 0
& \pminus 0   & \pminus 0   & \pminus 0   & \pminus 0   & \pminus 0   & \pminus 0   & \pminus 0   & \pminus 1
& \pminus 0   & \pminus 0   & \pminus 0   & \pminus 0   & \pminus 0   & \pminus 0   & \pminus 0
\\
  \pminus 0   & \pminus 0   & \pminus 0   & \pminus 0   & \pminus 0   & \pminus 0   & \pminus 0   & \pminus 0
& \pminus 0   & \pminus 0   & \pminus 0   & \pminus 0   & \pminus 0   & \pminus 0   & \pminus 0   & \pminus 0
& \pminus 0   & \pminus 0   & \pminus 0   & \pminus 0   & \pminus 0   & \pminus 0   & \pminus 0   & \pminus 0
& \pminus 0   & \pminus 1   & \pminus 0   & \pminus 0   & \pminus 0   & \pminus 0   & \pminus 0
\\
  \pminus 0   & \pminus 0   & \pminus 0   & \pminus 0   & \pminus 0   & \pminus 0   & \pminus 0   & \pminus 0
& \pminus 0   & \pminus 0   & \pminus 0   & \pminus 0   & \pminus 0   & \pminus 0   & \pminus 0   & \pminus 0
& \pminus 1   & \pminus 0   & \pminus 0   & \pminus 0   & \pminus 0   & \pminus 0   & \pminus 0   & \pminus 0
& \pminus 0   & \pminus 0   & \pminus 0   & \pminus 0   & \pminus 0   & \pminus 0   & \pminus 0
\\
  \pminus 0   & \pminus 0   & \pminus 0   & \pminus 0   & \pminus 0   & \pminus 0   & \pminus 0   & \pminus 0
& \pminus 0   & \pminus 0   & \pminus 0   & \pminus 0   & \pminus 0   & \pminus 0   & \pminus 0   & \pminus 1
& \pminus 0   & \pminus 0   & \pminus 0   & \pminus 0   & \pminus 0   & \pminus 0   & \pminus 0   & \pminus 0
& \pminus 0   & \pminus 0   & \pminus 0   & \pminus 0   & \pminus 0   & \pminus 0   & \pminus 0
\\
  \pminus 0   & \pminus 0   & \pminus 0   & \pminus 0   & \pminus 0   & \pminus 0   & \pminus 0   & \pminus 0
& \pminus 0   & \pminus 0   & \pminus 0   & \pminus 0   & \pminus 0   & \pminus 0   & \pminus 0   & \pminus 0
& \pminus 0   & \pminus 0   & \pminus 0   & \pminus 0   & \pminus 1   & \pminus 0   & \pminus 0   & \pminus 0
& \pminus 0   & \pminus 0   & \pminus 0   & \pminus 0   & \pminus 0   & \pminus 0   & \pminus 0
\\
  \pminus 0   & \pminus 0   & \pminus 0   & \pminus 0   & \pminus 0   & \pminus 0   & \pminus 0   & \pminus 0
& \pminus 0   & \pminus 0   & \pminus 0   & \pminus 0   & \pminus 0   & \pminus 0   & \pminus 0   & \pminus 0
& \pminus 0   & \pminus 0   & \pminus 0   & \pminus 1   & \pminus 0   & \pminus 0   & \pminus 0   & \pminus 0
& \pminus 0   & \pminus 0   & \pminus 0   & \pminus 0   & \pminus 0   & \pminus 0   & \pminus 0
\\
  \minus 1    & \minus \frac{3}{2}        & \pminus \frac{1}{2}       & \pminus 0   & \pminus 0   & \pminus 0
& \minus 1    & \pminus 0   & \pminus 0   & \pminus 0   & \pminus 0   & \pminus 0   & \pminus 2   & \pminus 1
& \pminus 0   & \pminus 0   & \pminus 0   & \pminus 0   & \pminus 0   & \pminus 0   & \pminus 0   & \pminus 0
& \pminus 0   & \pminus 0   & \pminus 0   & \pminus 0   & \pminus 0   & \pminus 0   & \pminus 0   & \pminus 0
& \pminus 0
\\
  \pminus 0   & \pminus 0   & \pminus 0   & \pminus 0   & \pminus 0   & \pminus 0   & \pminus 0   & \pminus 0
& \pminus 0   & \pminus 0   & \pminus 0   & \pminus 0   & \pminus 0   & \pminus 1   & \pminus 0   & \pminus 0
& \pminus 0   & \pminus 0   & \pminus 0   & \pminus 0   & \pminus 0   & \pminus 0   & \pminus 0   & \pminus 0
& \pminus 0   & \pminus 0   & \pminus 0   & \pminus 0   & \pminus 0   & \pminus 0   & \pminus 0
\\
  \pminus 0   & \pminus 0   & \pminus 0   & \pminus 0   & \pminus 0   & \pminus 0   & \pminus 0   & \pminus 0
& \pminus 0   & \pminus 0   & \pminus 0   & \pminus 0   & \pminus 1   & \pminus 0   & \pminus 0   & \pminus 0
& \pminus 0   & \pminus 0   & \pminus 0   & \pminus 0   & \pminus 0   & \pminus 0   & \pminus 0   & \pminus 0
& \pminus 0   & \pminus 0   & \pminus 0   & \pminus 0   & \pminus 0   & \pminus 0   & \pminus 0
\\
  \pminus 0   & \pminus 0   & \pminus 0   & \pminus 0   & \pminus 0   & \pminus 0   & \pminus 0   & \pminus 0
& \pminus 0   & \pminus 0   & \pminus 0   & \pminus 0   & \pminus 0   & \pminus 0   & \pminus 0   & \pminus 0
& \pminus 0   & \pminus 1   & \pminus 0   & \pminus 0   & \pminus 0   & \pminus 0   & \pminus 0   & \pminus 0
& \pminus 0   & \pminus 0   & \pminus 0   & \pminus 0   & \pminus 0   & \pminus 0   & \pminus 0
\\
  \pminus 0   & \pminus 0   & \pminus 0   & \pminus 0   & \pminus 0   & \pminus 0   & \pminus 0   & \pminus 0
& \pminus 0   & \pminus 0   & \pminus 0   & \pminus 0   & \pminus 0   & \pminus 0   & \pminus 1   & \pminus 0
& \pminus 0   & \pminus 0   & \pminus 0   & \pminus 0   & \pminus 0   & \pminus 0   & \pminus 0   & \pminus 0
& \pminus 0   & \pminus 0   & \pminus 0   & \pminus 0   & \pminus 0   & \pminus 0   & \pminus 0
\\
  \pminus 0   & \pminus 0   & \pminus 0   & \pminus 0   & \pminus \frac{3}{2}       & \pminus \frac{3}{2}
& \pminus 0   & \pminus 0   & \pminus 0   & \pminus 0   & \pminus 0   & \pminus 0   & \pminus 0   & \pminus 0
& \pminus 0   & \pminus 0   & \pminus 0   & \pminus 0   & \pminus 0   & \pminus 0   & \pminus 0   & \minus 6
& \pminus 0   & \pminus 0   & \pminus 0   & \pminus 0   & \pminus 0   & \pminus 0   & \pminus 0   & \pminus 0
& \pminus 0
\\
  \pminus 0   & \pminus 0   & \pminus 0   & \pminus 0   & \pminus 0   & \pminus 0   & \pminus 0   & \pminus 0
& \pminus 0   & \pminus 0   & \pminus 0   & \pminus 0   & \pminus 0   & \pminus 0   & \pminus 0   & \pminus 0
& \pminus 0   & \pminus 0   & \pminus 0   & \pminus 0   & \pminus 0   & \pminus 1   & \pminus 0   & \pminus 0
& \pminus 0   & \pminus 0   & \pminus 0   & \pminus 0   & \pminus 0   & \pminus 0   & \pminus 0
\\
  \pminus 0   & \pminus 0   & \pminus 0   & \pminus 0   & \pminus 0   & \pminus 0   & \pminus 0   & \pminus 0
& \pminus 0   & \pminus 0   & \pminus 0   & \pminus 0   & \pminus 0   & \pminus 0   & \pminus 0   & \pminus 0
& \pminus 0   & \pminus 0   & \pminus 0   & \pminus 0   & \pminus 0   & \pminus 0   & \pminus 1   & \pminus 0
& \pminus 0   & \pminus 0   & \pminus 0   & \pminus 0   & \pminus 0   & \pminus 0   & \pminus 0
\\
  \pminus 0   & \pminus 0   & \pminus 0   & \pminus 0   & \pminus 0   & \minus \frac{1}{6}        & \pminus 0
& \pminus 0   & \minus 1    & \pminus 0   & \pminus 0   & \minus \frac{1}{3}        & \pminus 0   & \pminus 0
& \pminus 0   & \pminus 0   & \pminus 0   & \pminus 0   & \pminus 0   & \pminus 0   & \pminus 0   & \pminus 0
& \pminus \frac{1}{3}       & \pminus 0   & \pminus 0   & \pminus 0   & \pminus 0   & \pminus 0   & \pminus 0
& \pminus 0   & \pminus 0
\\
  \pminus 0   & \pminus 0   & \pminus 0   & \pminus 0   & \pminus 0   & \pminus 0   & \pminus 0   & \pminus 0
& \pminus 0   & \pminus 0   & \pminus 0   & \pminus 0   & \pminus 0   & \pminus 0   & \pminus 0   & \pminus 0
& \pminus 0   & \pminus 0   & \pminus 1   & \pminus 0   & \pminus 0   & \pminus 0   & \pminus 0   & \pminus 0
& \pminus 0   & \pminus 0   & \pminus 0   & \pminus 0   & \pminus 0   & \pminus 0   & \pminus 0
\\
  \pminus 0   & \pminus 0   & \pminus 0   & \pminus 0   & \pminus 0   & \pminus 0   & \pminus 0   & \pminus 1
& \pminus 0   & \pminus 0   & \pminus 0   & \pminus 0   & \pminus 0   & \pminus 0   & \pminus 0   & \pminus 0
& \pminus 0   & \pminus 0   & \pminus 0   & \pminus 0   & \pminus 0   & \pminus 0   & \pminus 0   & \pminus 0
& \pminus 0   & \pminus 0   & \pminus 0   & \pminus 0   & \pminus 0   & \pminus 0   & \pminus 0
\\
  \pminus 0   & \pminus 0   & \pminus 0   & \pminus 0   & \pminus 0   & \pminus 0   & \pminus 0   & \pminus 1
& \pminus 0   & \pminus 0   & \pminus 0   & \pminus 0   & \pminus 0   & \pminus 0   & \pminus 0   & \pminus 0
& \pminus 0   & \pminus 0   & \pminus 0   & \pminus 0   & \pminus 0   & \pminus 0   & \pminus 0   & \pminus 0
& \pminus 0   & \pminus 0   & \pminus 0   & \pminus 0   & \pminus 0   & \pminus 0   & \pminus 0
\\
  \pminus 1   & \pminus 0   & \pminus 0   & \pminus 0   & \pminus 0   & \pminus 0   & \pminus 0   & \pminus 0
& \pminus 0   & \pminus 1   & \pminus 0   & \pminus 0   & \pminus 0   & \pminus 0   & \pminus 0   & \pminus 0
& \pminus 0   & \pminus 0   & \pminus 0   & \pminus 0   & \pminus 0   & \pminus 0   & \pminus 0   & \pminus 0
& \pminus 0   & \pminus 0   & \pminus 0   & \pminus 0   & \pminus 0   & \pminus 0   & \pminus 0
\\
  \pminus 0   & \pminus 0   & \pminus 0   & \pminus 0   & \pminus 0   & \pminus 0   & \pminus 1   & \pminus 0
& \pminus 0   & \pminus 0   & \pminus 0   & \pminus 0   & \pminus 0   & \pminus 0   & \pminus 0   & \pminus 0
& \pminus 0   & \pminus 0   & \pminus 0   & \pminus 0   & \pminus 0   & \pminus 0   & \pminus 0   & \pminus 0
& \pminus 0   & \pminus 0   & \pminus 0   & \pminus 0   & \pminus 0   & \pminus 0   & \pminus 0
\\
  \pminus 0   & \minus \frac{1}{2}        & \pminus 0   & \pminus 0   & \pminus 0   & \pminus 0   & \pminus 0
& \pminus 0   & \pminus 0   & \pminus 0   & \pminus 0   & \pminus 0   & \pminus 0   & \pminus 0   & \pminus 0
& \pminus 0   & \pminus 0   & \pminus 0   & \pminus 0   & \pminus 0   & \pminus 0   & \pminus 0   & \pminus 0
& \pminus 0   & \pminus 0   & \pminus 0   & \pminus 0   & \pminus 0   & \pminus 0   & \pminus 0   & \pminus 0
\\
  \pminus 0   & \pminus 0   & \pminus 0   & \pminus 0   & \pminus 0   & \pminus 0   & \pminus 0   & \pminus 0
& \pminus 0   & \pminus 0   & \pminus 0   & \pminus 1   & \pminus 0   & \pminus 0   & \pminus 0   & \pminus 0
& \pminus 0   & \pminus 0   & \pminus 0   & \pminus 0   & \pminus 0   & \pminus 0   & \pminus 0   & \pminus 0
& \pminus 0   & \pminus 0   & \pminus 0   & \pminus 0   & \pminus 0   & \pminus 0   & \pminus 0
\\
  \pminus 2   & \pminus 0   & \pminus 0   & \pminus 0   & \pminus 0   & \pminus 0   & \pminus 0   & \pminus 0
& \pminus 1   & \pminus 0   & \pminus 0   & \pminus 0   & \pminus 0   & \pminus 0   & \pminus 0   & \pminus 0
& \pminus 0   & \pminus 0   & \pminus 0   & \pminus 0   & \pminus 0   & \pminus 0   & \pminus 0   & \pminus 0
& \pminus 0   & \pminus 0   & \pminus 0   & \pminus 0   & \pminus 0   & \pminus 0   & \pminus 0
\\
  \pminus 0   & \pminus 0   & \pminus 0   & \pminus 0   & \pminus 0   & \pminus 0   & \pminus 0   & \pminus 0
& \pminus 0   & \pminus 0   & \minus 1    & \pminus 0   & \pminus 0   & \pminus 0   & \pminus 0   & \pminus 0
& \pminus 0   & \pminus 0   & \pminus 0   & \pminus 0   & \pminus 0   & \pminus 0   & \pminus 0   & \pminus 0
& \pminus 0   & \pminus 0   & \pminus 0   & \pminus 0   & \pminus 0   & \pminus 0   & \pminus 0
\\
  \pminus 0   & \pminus 0   & \pminus 0   & \pminus 1   & \pminus 0   & \pminus 0   & \pminus 0   & \pminus 0
& \pminus 0   & \pminus 0   & \pminus 0   & \pminus 0   & \pminus 0   & \pminus 0   & \pminus 0   & \pminus 0
& \pminus 0   & \pminus 0   & \pminus 0   & \pminus 0   & \pminus 0   & \pminus 0   & \pminus 0   & \pminus 0
& \pminus 0   & \pminus 0   & \pminus 0   & \pminus 0   & \pminus 0   & \pminus 0   & \pminus 0
\\
\end{matrix}
~\right)
}
\label{eq:K-matrix}
\end{equation}
\begin{table}[htbp]
\centering
\renewcommand\arraystretch{1.2}
\begin{tabular}{
p{0.8cm}<{\centering}
l}
\toprule[1.5pt]
	  $\textbf{G}$
   &
      {\it Analytic} $\big[\mathcal{O}(z\log^n z)\big]$
      ~\big/~
      {\it Analytic} $\big[\mathcal{O}(z^2\log^n z)\big]$
      ~\big/~
      \texttt{pySecDec}
      \\
\midrule[1pt]
      \multirow{3}*{$g_{13}$}
   &  {\small
      $41.2060921304982\,\epsilon^2
     + 363.015345460615\,\epsilon^3
     + 1991.43341346475\,\epsilon^4$
      } \\
   &  {\small
      $41.2060921304982\,\epsilon^2
     + 363.015343632027\,\epsilon^3
     + 1991.43340188216\,\epsilon^4$
      } \\
   &  {\small
      $41.20609213041(25)\,\epsilon^2
     + 363.01537(5)\,\epsilon^3
     + 1991.4338(8)\,\epsilon^4$
      } \\
\midrule[0.8pt]
      \multirow{3}*{$g_{26}$}
   &  {\small
      $47.5646588551249\,\epsilon^3
     + 359.901210192870\,\epsilon^4$
      } \\
   &  {\small
      $47.5646586479934\,\epsilon^3
     + 359.901217045946\,\epsilon^4$
      } \\
   &  {\small
      $47.564661(5)\,\epsilon^3
     + 359.901218(29)\,\epsilon^4$
      } \\
\midrule[0.8pt]
      \multirow{3}*{$g_{41}$}
   &  {\small
      $-41.1936546266871\,\epsilon^3
     - 312.411476285700\,\epsilon^4$
      } \\
   &  {\small
      $-41.1936549977326\,\epsilon^3
     - 312.411485581619\,\epsilon^4$
      } \\
   &  {\small
      $-41.193654989(13)\,\epsilon^3
     - 312.41134(22)\,\epsilon^4$
      } \\
\midrule[0.8pt]
      \multirow{3}*{$g_{46}$}
   &  {\small
      $7.71826798162703\,\epsilon^2
     + 22.3660080321417\,\epsilon^3
     - 81.8728881227263\,\epsilon^4$
      } \\
   &  {\small
      $7.71826825644108\,\epsilon^2
     + 22.3660072472345\,\epsilon^3
     - 81.8729064820082\,\epsilon^4$
      } \\
   &  {\small
      $7.718268257(1)\,\epsilon^2
     + 22.3647(16)\,\epsilon^3
     - 81.885(12)\,\epsilon^4$
      } \\
\bottomrule[1.5pt]
\end{tabular}
\caption{
Comparison between the numerical results obtained from our analytic expressions and \texttt{pySecDec} for $g_{13}$, $g_{26}$, $g_{41}$ and $g_{46}$ at $(s,\, t,\, m_{\ell}^2,\, m_W^2) = (-5,\, -2,\, 10^{-3},\, 4)~ \text{GeV}^2$.}
\label{tab1}
\end{table}

\section{Summary}
\label{sec:4}
\par
The DY production of dilepton via $Z$ and single lepton via $W$ in hadron-hadron collisions is one of the cutting-edge topics in the field of high energy physics. In this paper we present an analytic calculation of the MIs for the mixed QCD-QED two-loop virtual corrections to the charged-current DY process $q\bar{q}^{\prime} \rightarrow \ell \nu$. The scalar Feynman integrals involved in the non-factorizable corrections can be expressed in terms of the linear combinations of $46$ MIs, which satisfy a set of first-order differential equations. By means of a suitable basis transformation, we obtain a canonical set of MIs. The $\epsilon$-dependence of the canonical differential system is factorized from the kinematics. Therefore, the canonical MIs, as the solution of the canonical differential system, can be cast in a Taylor series around $d = 4$, with coefficients written in terms of Chen's iterated integrals. After an expansion for small $z$, all the entries of the coefficient matrices are simple rational functions, and thus the Chen's iterated integrals can be expressed in terms of GPLs. The boundary constants of these canonical MIs are fixed by using known or simpler integrals as independent input and by requiring the regularity of the solution at some special kinematic limits. We solve the canonical differential system in the second-order approximation and then obtain the Taylor series expansions of the $46$ canonical MIs around $d = 4$ up to $\mathcal{O}(\epsilon^4)$, which contain all the logarithmic terms of the form $z^{m}\log^n z~ (m = 0, 1,~ n = 0, 1, 2, 3, 4)$. As for the factorizable corrections, the related MIs are either included in the $46$ MIs for the non-factorizable corrections or obtained from the literature. Our analytic results can help to implement a flexible and efficient MC program for precision analysis of charged-current DY processes when the finite-lepton-mass effect is taken into consideration.

\section*{Acknowledgments}
This work is supported in part by the National Natural Science Foundation of China (Grants No. 11775211 and No. 12061141005) and the CAS Center for Excellence in Particle Physics (CCEPP).

\appendix

\section{Mixed QCD-QED virtual corrections to $q\bar{q}^{\prime} \rightarrow W^{\ast}$}
\label{sec:appendix-A}
\par
\begin{figure}[htbp]
\centering
\includegraphics[width=0.9\textwidth]{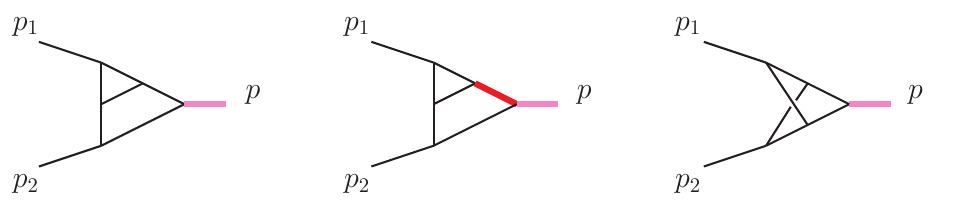}
\caption{Three representative topologies of Feynman diagrams belonging to $\mathcal{F}_1$, $\mathcal{F}_2$ and $\mathcal{F}_3$ respectively for mixed QCD-QED two-loop virtual corrections to $q\bar{q}^{\prime} \rightarrow W^{\ast}$.}
\label{fig3}
\end{figure}
\begin{figure}[H]
\centering
\includegraphics[width=0.9\textwidth]{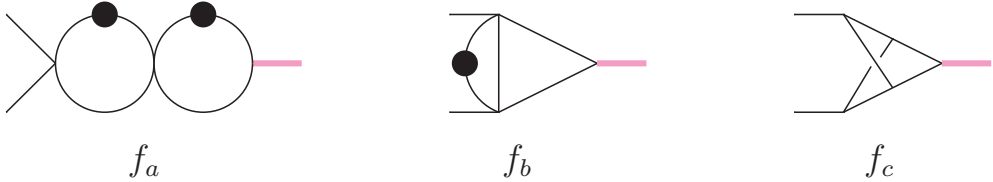}
\caption{Three MIs for mixed QCD-QED two-loop virtual corrections to $q\bar{q}^{\prime} \rightarrow W^{\ast}$.}
\label{fig4}
\end{figure}

The $20$ two-loop Feynman diagrams for the mixed QCD-QED virtual corrections to $q\bar{q}^{\prime} \rightarrow W^{\ast}$ can be categorized into $8$ topologies belonging to $3$ vertex integral families $\mathcal{F}_{1,2,3}$. Some representative topologies of these Feynman diagrams are depicted in Figure \ref{fig3}.
\begin{itemize}
\item {\it Family $\mathcal{F}_1$}:
      \begin{equation}
      \begin{aligned}
      D_1 &= l_1^2\,, \qquad &
      D_2 &= (l_1+p_1)^2\,, \qquad &
      D_3 &= (l_1-p_2)^2\,, \qquad \\
      D_4 &= l_2^2\,, \qquad &
      D_5 &= (l_2+p_1)^2\,, \qquad &
      D_6 &= (l_2-p_2)^2\,, \qquad \\
      D_7 &= (l_1-l_2)^2
      \end{aligned}
      \end{equation}
      {\it $3$ topologies $\subset \mathcal{F}_1$}:
      \begin{equation}
      [1, 1, 1, 1, 1, 0, 1]\,,
      \quad
      [1, 1, 1, 0, 1, 1, 1]\,,
      \quad
      [1, 1, 1, 0, 1, 0, 1]
      \label{eq:top-F1}
      \end{equation}
      It is worth mentioning that $[1, 1, 1, 0, 1, 0, 1]$ is the sub-sector of $[1, 1, 1, 1, 1, 0, 1]$ and $[1, 1, 1, 0, 1, 1, 1]$. The two-loop scalar integrals belonging to $\mathcal{S}_1$,
      \begin{equation}
      \mathcal{S}_1
      \,\equiv\,
      [1, 1, 1, 1, 1, 0, 1]_{\digamma}
      \cup \,
      [1, 1, 1, 0, 1, 1, 1]_{\digamma}\,,
      \end{equation}
      can be reduced to $3$ MIs $f_5$, $f_a$ and $f_b$, where $f_a$ and $f_b$ are depicted schematically in Figure \ref{fig4}.
\item {\it Family $\mathcal{F}_2$}:
      \begin{equation}
      \begin{aligned}
      D_1 &= l_1^2\,, \qquad &
      D_2 &= (l_1+p_1)^2-m_W^2\,, \qquad &
      D_3 &= (l_1-p_2)^2\,, \qquad \\
      D_4 &= l_2^2\,, \qquad &
      D_5 &= (l_2+p_1)^2\,, \qquad &
      D_6 &= (l_2-p_2)^2\,, \qquad \\
      D_7 &= (l_1-l_2)^2
      \end{aligned}
      \end{equation}
      {\it $4$ topologies $\subset \mathcal{F}_2$}:
      \begin{equation}
      [1, 1, 1, 1, 1, 0, 1]\,,
      \quad
      [1, 1, 1, 1, 0, 1, 1]\,,
      \quad
      [0, 1, 1, 1, 1, 1, 1]\,,
      \quad
      [1, 1, 1, 1, 0, 0, 1]
      \label{eq:top-F2}
      \end{equation}
      In Eq.(\ref{eq:top-F2}) the last sector is the sub-sector of the first two sectors. All the two-loop scalar integrals of $\mathcal{F}_2$ involved in the mixed QCD-QED virtual corrections to $q\bar{q}^{\prime} \rightarrow W^{\ast}$ belong to $\mathcal{S}_2$,
      \begin{equation}
      \mathcal{S}_2
      \,\equiv\,
      [1, 1, 1, 1, 1, 0, 1]_{\digamma}
      \cup \,
      [1, 1, 1, 1, 0, 1, 1]_{\digamma}
      \cup \,
      [0, 1, 1, 1, 1, 1, 1]_{\digamma}\,.
      \end{equation}
      The $9$ MIs of this integral set can be chosen as $f_{2}, f_{3}, f_{5}, f_{9}, f_{11}, f_{12}, f_{14}, f_{15}, f_{28}$.
\item {\it Family $\mathcal{F}_3$}:
      \begin{equation}
      \begin{aligned}
      D_1 &= l_1^2\,, \qquad &
      D_2 &= (l_1+p_1)^2\,, \qquad &
      D_3 &= (l_1-p_2)^2\,, \qquad \\
      D_4 &= (l_2+p_1)^2\,, \qquad &
      D_5 &= (l_2-p_2)^2\,, \qquad &
      D_6 &= (l_1-l_2)^2\,, \qquad \\
      D_7 &= (l_1-l_2+p_2)^2
      \end{aligned}
      \end{equation}
      {\it $1$ topology $\subset \mathcal{F}_3$}:
      \begin{equation}
      [1, 1, 0, 1, 1, 1, 1]
      \label{eq:top-F3}
      \end{equation}
      Unlike the $7$ topologies in Eqs.(\ref{eq:top-F1}) and (\ref{eq:top-F2}), $[1, 1, 0, 1, 1, 1, 1]$ in Eq.(\ref{eq:top-F3}) is a three-point non-planar topology, which is represented schematically by the non-planar graph in Figure \ref{fig3}. The two-loop scalar integrals induced by the non-planar Feynman diagrams via tensor reduction belong to $\mathcal{S}_3$,
      \begin{equation}
      \mathcal{S}_3
      \,\equiv\,
      [1, 1, 0, 1, 1, 1, 1]_{\digamma}\,,
      \end{equation}
      and can be reduced to a set of $3$ MIs $f_5$, $f_b$ and $f_c$, where $f_c$ is a non-planar two-loop scalar integral defined by the last graph in Figure \ref{fig4}.
\end{itemize}

\par
In summary, there are totally $12$ MIs for the mixed QCD-QED two-loop virtual corrections to $q\bar{q}^{\prime} \rightarrow W^{\ast}$, which can be chosen as
\begin{equation}
g_{2},~ g_{3},~ g_{5},~ g_{9},~ g_{11},~ g_{12},~ g_{14},~ g_{15},~ g_{28},~ f_{a},~ f_{b},~ f_{c}\,.
\label{eq:12MIs}
\end{equation}
All the two-loop scalar integrals involved in the calculation of mixed QCD-QED virtual corrections to $q\bar{q}^{\prime} \rightarrow W^{\ast}$ contain at most two energy scales, the invariant mass $\sqrt{s}$ and the $W$-boson mass $m_W$. Being the same as $f_i$ and $g_i~ (i = 1, ..., 46)$, $f_{a, b, c}$ have been nondimensionalized by the $W$-boson mass; thus the $12$ MIs in Eq.(\ref{eq:12MIs}) should depend only on the dimensionless variable $x = -s/m_W^2$\footnote{From Figure \ref{fig2} and Eq.(\ref{eq:UTbasis}), we may conclude that $g_{2,3,5,9,11,12,14,15,28}$ are only functions of $x$, while the other $37$ components of $\textbf{G}$ depend also on $y$ and/or $z$.}. The first $9$ MIs, $g_{2,3,5,9,11,12,14,15,28}$, have been included in the $46$ MIs for the non-factorizable mixed QCD-QED two-loop virtual corrections to $q\bar{q}^{\prime} \rightarrow \ell \nu$. For the remaining $3$ MIs, $f_{a, b, c}$ (depicted in Figure \ref{fig4}), their analytic expressions in terms of GPLs up to weight $4$ can be obtained from the literature \cite{Bonciani:2016ypc,Gehrmann:2014bfa}.

\section{Auxiliary integrals}
\label{sec:appendix-B}
\par
\begin{figure}[H]
\centering
\includegraphics[width=0.4\textwidth]{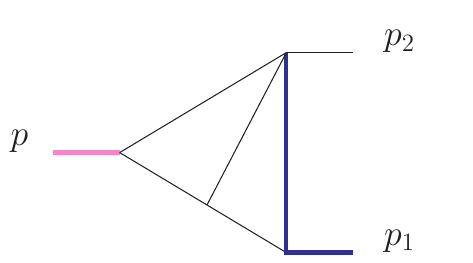}
\caption{
Sector $[1, 1, 1, 1, 1, 0, 0]$ of the vertex integral family identified by Eq.(\ref{eq:family-V}).}
\label{fig5}
\end{figure}
\begin{figure}[H]
\centering
\includegraphics[width=1.0\textwidth]{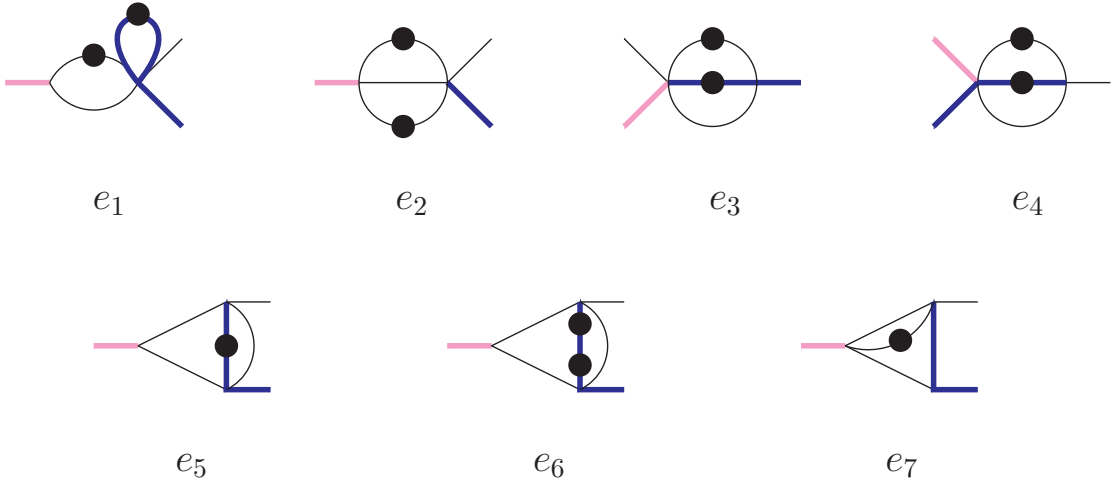}
\caption{
A set of pre-canonical MIs for $\mathcal{S}_{\text{aux}}$.}
\label{fig6}
\end{figure}

In this appendix we discuss the computation of the auxiliary vertex integrals used to extract the boundary constants of $g_{13}$ and $g_{18}$ in Section \ref{subsec:3-3}. The two-loop vertex integral family considered here is identified by the set of propagators
\begin{equation}
\begin{aligned}
	D_1 &= l_1^2\,, \qquad &
	D_2 &= l_2^2\,, \qquad &
	D_3 &= (l_1-l_2)^2\,, \qquad \\
	D_4 &= (l_1+p_1)^2-m_{\ell}^2\,, \qquad &
	D_5 &= (l_2+p_1+p_2)^2\,, \qquad &
	D_6 &= (l_2+p_1)^2\,, \qquad \\
	D_7 &= (l_1+p_1+p_2)^2\,,
\label{eq:family-V}
\end{aligned}
\end{equation}
with external momenta $p_1$, $p_2$ and $p$ satisfying
\begin{equation}
	p_1^2=m_{\ell}^2\,, \qquad p_2^2=0\,, \qquad (p_1+p_2)^2 = p^2 =s\,.
\end{equation}
Now we focus on the sector $[1, 1, 1, 1, 1, 0, 0]$ (depicted in Figure \ref{fig5}) of this integral family. It is obvious that
\begin{equation}
\begin{aligned}
f_{13} & ~\in~ [1, 0, 1, 1, 1, 0, 0]
\\
f_{18} & ~\in~ [0, 1, 1, 1, 1, 0, 0]
\end{aligned}
~\,\subset~
[1, 1, 1, 1, 1, 0, 0]_{\digamma}\,.
\end{equation}
All the integrals belonging to $\mathcal{S}_{\text{aux}} \equiv [1, 1, 1, 1, 1, 0, 0]_{\digamma}$ have the form of
\begin{equation}
\int \mathcal{D}^d l_1 \mathcal{D}^d l_2
\frac{D_6^{n_6} D_7^{n_7}}{D_1^{n_1}D_2^{n_2}D_3^{n_3}D_4^{n_4}D_5^{n_5}}\,,
\qquad
(n_{1, ..., 5} \in \mathbb{Z},\quad n_{6,7} \in \mathbb{N})\,,
\label{set-Saux}
\end{equation}
and can be reduced to a set of $7$ MIs shown in Figure \ref{fig6}. These MIs ($e_{i},~ i = 1, ..., 7$) have been nondimensionalized by $(-s)$, and thus can be expressed as functions of the dimensionless variable $\kappa$,
\begin{equation}
\kappa = -\frac{m_{\ell}^2}{s}\,.
\end{equation}
The canonical basis $\boldsymbol{\Gamma} = (\gamma_1, \gamma_2, ..., \gamma_7)^{T}$, defined by Eq.(\ref{eq:UT-appendix}),
\begin{align}
\gamma_1 & = \epsilon^2\,e_1\,,
         &
\gamma_2 & = \epsilon^2\,e_2\,, \nonumber \\
\gamma_3 & = \epsilon^2\,e_3\,\kappa\,,
         &
\gamma_4 & = \epsilon\,(1-\epsilon)\,e_4\,\kappa\,,
\label{eq:UT-appendix}
\\
\gamma_5 & = \epsilon^3\,e_5\,(1+\kappa)\,,
         &
\gamma_6 & = \epsilon^2\,e_6\,\kappa\,(1+\kappa)\,, \nonumber \\
\gamma_7 & = \epsilon^3\,e_7\,(1+\kappa)\,,
         &
         & \nonumber
\end{align}
satisfies the following $d\log$-form total differential equation,
\begin{equation}
d\boldsymbol{\Gamma}
=
\epsilon\,
\Big[
\mathbb{M}_1\, d\log(\kappa) + \mathbb{M}_{2}\, d\log(\kappa+1)
\Big]
\,
\boldsymbol{\Gamma}\,,
\label{eq:auxdlog}
\end{equation}
where
\begin{equation}
\mathbb{M}_1
=
\left(
\begin{matrix}
 \minus 1   & \pminus 0  & \pminus 0  & \pminus 0  & \pminus 0  & \pminus 0  & \pminus 0 \\
 \pminus 0  & \pminus 0  & \pminus 0  & \pminus 0  & \pminus 0  & \pminus 0  & \pminus 0 \\
 \pminus 0  & \pminus 0  & \minus 2   & \pminus 0  & \pminus 0  & \pminus 0  & \pminus 0 \\
 \pminus 0  & \pminus 0  & \pminus 0  & \minus 2   & \pminus 0  & \pminus 0  & \pminus 0 \\
 \pminus 0  & \pminus 0  & \pminus 2  & \pminus \frac{1}{2}     & \pminus 0  & \pminus 2  & \pminus 0 \\
 \pminus \frac{1}{2}     & \pminus 0  & \pminus 0  & \minus \frac{1}{2}      & \pminus 0  & \minus 2  & \pminus 0 \\
 \pminus 0  & \minus \frac{1}{2}      & \pminus 0  & \minus \frac{1}{2}      & \pminus 0  & \pminus 0 & \minus 1 \\
\end{matrix}
~\right)\,, \qquad
\mathbb{M}_{2}
=
\left(
\begin{matrix}
 \pminus 0  & \pminus 0  & \pminus 0  & \pminus 0  & \pminus 0  & \pminus 0  & \pminus 0 \\
 \pminus 0  & \pminus 0  & \pminus 0  & \pminus 0  & \pminus 0  & \pminus 0  & \pminus 0 \\
 \pminus 0  & \pminus 0  & \pminus 0  & \pminus 0  & \pminus 0  & \pminus 0  & \pminus 0 \\
 \pminus 0  & \pminus 0  & \pminus 0  & \pminus 0  & \pminus 0  & \pminus 0  & \pminus 0 \\
 \pminus 0  & \pminus 0  & \pminus 0  & \pminus 0  & \minus 1   & \minus 2   & \pminus 0 \\
 \pminus 0  & \pminus 0  & \pminus 0  & \pminus 0  & \pminus 3  & \pminus 4  & \pminus 0 \\
 \pminus 0  & \pminus 0  & \pminus 0  & \pminus 0  & \pminus 0  & \pminus 0  & \pminus 2 \\
\end{matrix}
~\right)\,.
\end{equation}
The same strategy as in Section \ref{subsec:3-2} is applied here to solve the canonical differential system in Eq.(\ref{eq:auxdlog}). To determine the unknown integration constants, the four simple integrals $\gamma_{1,2,3,4}$ serve as input to this system \cite{Bonciani:2016ypc,Hasan:2020vwn}. As for $\gamma_{5,6,7}$, the integration constants are fixed by the regularity at $\kappa \rightarrow -1$. Then we obtain the solution of the canonical differential system (\ref{eq:auxdlog}) in terms of GPLs of argument $\kappa$\footnote{We have also calculated $\boldsymbol{\Gamma}$ by using the \texttt{Maple} program \texttt{HyperInt} \cite{Panzer:2014caa}, and obtained the same results as in Eq.(\ref{eq:MIs-Gamma}).},
{
\small
\setlength{\lineskip}{4.5pt}
\setlength{\lineskiplimit}{6pt}
\begin{align}
\gamma_{1}=\,
          & 1 - \epsilon\, G(0;\kappa)
	        + \epsilon^2 \big[ G(0,0;\kappa)
			                 + \zeta(2) \big]
	        - \epsilon^3 \big[ \zeta(2)\, G(0;\kappa)
			                 + G(0,0,0;\kappa)
			                 - 2\, \zeta(3) \big] \nn
          & - \epsilon^4/4 \big[ 8\, \zeta(3)\, G(0;\kappa)
                               - 4\, \zeta(2)\, G(0,0;\kappa)
			                   - 4\, G(0,0,0,0;\kappa)
			                   - 19\, \zeta(4) \big]
            + \mathcal{O}(\epsilon^5)\,, \nn
\gamma_{2}=\,
          & - 1
	        + 6\, \epsilon^3\, \zeta(3)
	        + 9\, \epsilon^4\, \zeta(4)
            + \mathcal{O}(\epsilon^5)\,, \nn
\gamma_{3}=\,
          & - 1/4
	        + \epsilon/2\, G(0;\kappa)
	        - \epsilon^2/2 \big[ 2\, G(0,0;\kappa)
                               + 3\, \zeta(2) \big]
	        + \epsilon^3 \big[ 3\, \zeta(2) \,G(0;\kappa)
			                 + 2\, G(0,0,0;\kappa)
                             - 3\, \zeta(3) \big] \nn
	      & + 2\, \epsilon^4 \big[ 3\,\zeta(3)\,G(0;\kappa)
                                 - 3\,\zeta(2)\,G(0,0;\kappa)
                                 - 2\,G(0,0,0,0;\kappa)
			                     - 12\,\zeta(4)
                                 \big]
            + \mathcal{O}(\epsilon^5)\,, \nn
\gamma_{4}=\,
          & 1
	        - 2\, \epsilon\, G(0;\kappa)
	        + 4\, \epsilon^2 \big[ G(0,0;\kappa)
			                     + \zeta(2) \big]
	        - 2\, \epsilon^3 \big[ 4\, \zeta(2)\, G(0;\kappa)
			                     + 4\, G(0,0,0;\kappa)
			                     - \zeta(3) \big] \nn
          & - \epsilon^4 \big[ 4\, \zeta(3)\, G(0;\kappa)
                             - 16\, \zeta(2)\, G(0,0;\kappa)
			                 - 16\, G(0,0,0,0;\kappa)
			                 - 31\, \zeta(4) \big]
            + \mathcal{O}(\epsilon^5)\,, \nn
\gamma_{5}=\,
          & \epsilon^3 \big[ 2\, \zeta(2)\, G(0;\kappa)
                           - 3\, \zeta(2)\, G(-1;\kappa)
                           + G(0,0,0;\kappa)
			               - G(-1,0,0;\kappa)
			               - \zeta(3) \big] \nn
          & - \epsilon^4 \big[ \zeta(3)\, G(0;\kappa)
                             + 3\, \zeta(3)\, G(-1;\kappa)
                             + 7\, \zeta(2)\, G(0,0;\kappa)
                             - 7\, \zeta(2)\, G(-1,0;\kappa) \nn
          &                  - 12\, \zeta(2)\, G(0,-1;\kappa)
                             + 9\, \zeta(2)\, G(-1,-1;\kappa)
		                     + 5\, G(0,0,0,0;\kappa)
                             - 4\, G(-1,0,0,0;\kappa) \nn
          &                  - 4\, G(0,-1,0,0;\kappa)
		                     + 3\, G(-1,-1,0,0;\kappa)
			                 + 14\, \zeta(4) \big]
            + \mathcal{O}(\epsilon^5)\,, \nn
\gamma_{6}=\,
          & \epsilon^2/2 \big[ G(0,0;\kappa)
			                 + 3\, \zeta(2) \big]
	        - \epsilon^3/2 \big[ 9\, \zeta(2)\, G(0;\kappa)
                               - 12\, \zeta(2)\, G(-1;\kappa)
                               + 5\, G(0,0,0;\kappa) \nn
          &                    - 4\, G(-1,0,0;\kappa)
                               - 4\, \zeta(3) \big]
            - \epsilon^4/8 \big[ 32\, \zeta(3)\, G(0;\kappa)
                               - 40\, \zeta(3)\, G(-1;\kappa)
                               - 100\, \zeta(2)\, G(0,0;\kappa) \nn
          &                    + 96\, \zeta(2)\, G(-1,0;\kappa)
                               + 96\, \zeta(2)\, G(0,-1;\kappa)
                               - 120\, \zeta(2)\, G(-1,-1;\kappa)
                               - 68\, G(0,0,0,0;\kappa) \nn
          &                    + 56\, G(-1,0,0,0;\kappa)
                               + 32\, G(0,-1,0,0;\kappa)
                               - 40\, G(-1,-1,0,0;\kappa)
                               - 285\, \zeta(4)
                               \big]
            + \mathcal{O}(\epsilon^5)\,, \nn
\gamma_{7}=\,
          & \epsilon^2 \big[ G(0,0;\kappa)
			               + 3\, \zeta(2) \big]
	        - \epsilon^3 \big[ 5\, \zeta(2)\, G(0;\kappa)
                             - 6\, \zeta(2)\, G(-1;\kappa)
                             + 3\, G(0,0,0;\kappa)
                             - 2\, G(-1,0,0;\kappa) \nn
		  &                  - 2\, \zeta(3) \big]
            - \epsilon^4/4 \big[ 24\, \zeta(3)\, G(0;\kappa)
                               - 16\, \zeta(3)\, G(-1;\kappa)
                               - 36\, \zeta(2)\, G(0,0;\kappa)
                               + 40\, \zeta(2)\, G(-1,0;\kappa) \nn
          &                    + 24\, \zeta(2)\, G(0,-1;\kappa)
                               - 48\, \zeta(2)\, G(-1,-1;\kappa)
                               - 28\, G(0,0,0,0;\kappa)
                               + 24\, G(-1,0,0,0;\kappa) \nn
          &                    + 8\, G(0,-1,0,0;\kappa)
                               - 16\, G(-1,-1,0,0;\kappa)
                               - 89\, \zeta(4) \big]
            + \mathcal{O}(\epsilon^5)\,.
\label{eq:MIs-Gamma}
\end{align}
}
By the definition of $\textbf{G}$ and $\boldsymbol{\Gamma}$ we can see that
\begin{equation}
(g_{1},\, g_{5},\, g_{10},\, g_{6},\, g_{18},\, g_{19},\, g_{13})^{T}
=
x^{-2\epsilon}\,
\boldsymbol{\Gamma}
\quad
\text{with}
\quad
\kappa = z/x\,.
\label{eq:G-Gamma}
\end{equation}
Thus, the boundary constants of $g_{13}$ and $g_{18}$ are given by the constant components of $\gamma_{7}$ and $\gamma_5$, respectively,
\begin{equation}
\begin{aligned}
c_{13} &= \gamma_{7,\text{const}}
        = 3\, \zeta(2)\, \epsilon^2
        + 2\, \zeta(3)\, \epsilon^3
        + \frac{89}{4}\, \zeta(4)\, \epsilon^4
        + \mathcal{O}(\epsilon^5)\,, \\
c_{18} &= \gamma_{5,\text{const}}
        = - \zeta(3)\, \epsilon^3
        -14\, \zeta(4)\, \epsilon^4
        + \mathcal{O}(\epsilon^5)\,.
\end{aligned}
\end{equation}

\section{Asymptotic behaviour of $\textbf{G}$}
\label{sec:appendix-C}
\par
We analyze the asymptotic behaviour of $\textbf{G}$ in the limit $\alpha_{i} \rightarrow 0~ (i = 1, ..., 7)$ by using the code \texttt{asy.m} \cite{Pak:2010pt,Jantzen:2012mw} included in \texttt{FIESTA} \cite{Smirnov:2015mct,Smirnov:2021rhf} and summarize it in Table \ref{table:limit}.
\begin{longtable}{
c
p{1.5cm}<{\centering}
p{1.5cm}<{\centering}
p{1.5cm}<{\centering}
p{1.5cm}<{\centering}
p{1.5cm}<{\centering}
p{1.8cm}<{\centering}
p{2.5cm}<{\centering}
}
\toprule[1.5pt]
$g_i$   &  $s \rightarrow 0$  &  $s \rightarrow -t$          &  $s \rightarrow m_W^2$
        &  $t \rightarrow 0$  &  $m_{\ell}^2 \rightarrow 0$  &  $t \rightarrow -m_W^2$
        &  \makecell*{$\displaystyle s \rightarrow \dfrac{m_W^2 t}{t-m_W^2}$}  \\
\midrule[0.8pt]
\endfirsthead
\midrule[1.5pt]
\endfoot
\midrule[1.5pt]
$g_i$   &  $s \rightarrow 0$  &  $s \rightarrow -t$          &  $s \rightarrow m_W^2$
        &  $t \rightarrow 0$  &  $m_{\ell}^2 \rightarrow 0$  &  $t \rightarrow -m_W^2$
        &  \makecell*{$\displaystyle s \rightarrow \dfrac{m_W^2 t}{t-m_W^2}$} \\
\midrule[0.8pt]
\endhead
\caption{The kinematic limits $s \rightarrow 0$, $s \rightarrow m_W^2$, $s \rightarrow -t$, $s \rightarrow m_W^2t/(t-m_W^2)$, $t \rightarrow 0$, $t \rightarrow -m_W^2$ and $m_{\ell}^2 \rightarrow 0$ correspond to $\alpha_{i} \rightarrow 0,~ i = 1, ..., 7$, respectively. The chechmark means the MI is analytic, while the cross singular, in the corresponding limit.} \\
\endlastfoot
\label{table:limit}
$ 1$  & $\times$    & \checkmark  & \checkmark  & \checkmark  & $\times$    & \checkmark  & \checkmark  \\
$ 2$  & $\times$    & \checkmark  & \checkmark  & \checkmark  & \checkmark  & \checkmark  & \checkmark  \\
$ 3$  & \checkmark  & \checkmark  & $\times$    & \checkmark  & \checkmark  & \checkmark  & \checkmark  \\
$ 4$  & \checkmark  & \checkmark  & $\times$    & \checkmark  & $\times$    & \checkmark  & \checkmark  \\
$ 5$  & $\times$    & \checkmark  & \checkmark  & \checkmark  & \checkmark  & \checkmark  & \checkmark  \\
$ 6$  & \checkmark  & \checkmark  & \checkmark  & \checkmark  & $\times$    & \checkmark  & \checkmark  \\
$ 7$  & \checkmark  & \checkmark  & \checkmark  & $\times$    & $\times$    & \checkmark  & \checkmark  \\
$ 8$  & \checkmark  & \checkmark  & \checkmark  & $\times$    & \checkmark  & \checkmark  & \checkmark  \\
\end{longtable}

\section{Explicit expressions of $\textbf{G}$}
\label{sec:app3}
The analytic expressions of the $46$ canonical MIs $g_{i}~ (i=1, ..., 46)$ in terms of GPLs up to $\mathcal{O}(\epsilon^3)$ in the lowest-order approximation are listed as follows:
{
\small
\setlength{\lineskip}{4.5pt}
\setlength{\lineskiplimit}{6pt}
%
}

\bibliographystyle{JHEP}
\bibliography{references}

\providecommand{\href}[2]{#2}\begingroup\raggedright\begin{thebibliography}{10}

\bibitem{Drell:1970wh}
S.~D. Drell and T.-M. Yan, \emph{{Massive lepton pair production in
  hadron-hadron collisions at high-energies}},
  \href{https://doi.org/10.1103/PhysRevLett.25.316}{\emph{Phys. Rev. Lett.}
  {\bfseries 25} (1970) 316}.

\bibitem{Aaboud:2016btc}
{\scshape ATLAS} collaboration, \emph{{Precision measurement and interpretation
  of inclusive $W^+$, $W^-$ and $Z/\gamma^*$ production cross sections with the
  ATLAS detector}},
  \href{https://doi.org/10.1140/epjc/s10052-017-4911-9}{\emph{Eur. Phys. J. C}
  {\bfseries 77} (2017) 367}
  [\href{https://arxiv.org/abs/1612.03016}{{\ttfamily 1612.03016}}].

\bibitem{Aaboud:2017svj}
{\scshape ATLAS} collaboration, \emph{{Measurement of the $W$-boson mass in pp
  collisions at $\sqrt{s}=7~ \text{TeV}$ with the ATLAS detector}},
  \href{https://doi.org/10.1140/epjc/s10052-017-5475-4}{\emph{Eur. Phys. J. C}
  {\bfseries 78} (2018) 110}
  [\href{https://arxiv.org/abs/1701.07240}{{\ttfamily 1701.07240}}].

\bibitem{Camarda:2016twt}
S.~Camarda, J.~Cuth and M.~Schott, \emph{{Determination of the muonic branching
  ratio of the $W$ boson and its total width via cross-section measurements at
  the Tevatron and LHC}},
  \href{https://doi.org/10.1140/epjc/s10052-016-4461-6}{\emph{Eur. Phys. J. C}
  {\bfseries 76} (2016) 613}
  [\href{https://arxiv.org/abs/1607.05084}{{\ttfamily 1607.05084}}].

\bibitem{Aad:2015uau}
{\scshape ATLAS} collaboration, \emph{{Measurement of the forward-backward
  asymmetry of electron and muon pair-production in $pp$ collisions at
  $\sqrt{s} = 7~ \text{TeV}$ with the ATLAS detector}},
  \href{https://doi.org/10.1007/JHEP09(2015)049}{\emph{JHEP} {\bfseries 09}
  (2015) 049} [\href{https://arxiv.org/abs/1503.03709}{{\ttfamily
  1503.03709}}].

\bibitem{Sirunyan:2018swq}
{\scshape CMS} collaboration, \emph{{Measurement of the weak mixing angle using
  the forward-backward asymmetry of Drell-Yan events in $pp$ collisions at $8~
  \text{TeV}$}},
  \href{https://doi.org/10.1140/epjc/s10052-018-6148-7}{\emph{Eur. Phys. J. C}
  {\bfseries 78} (2018) 701}
  [\href{https://arxiv.org/abs/1806.00863}{{\ttfamily 1806.00863}}].

\bibitem{Khachatryan:2016pev}
{\scshape CMS} collaboration, \emph{{Measurement of the differential cross
  section and charge asymmetry for inclusive $pp \rightarrow W^{\pm} + X$
  production at $\sqrt{s} = 8~ \text{TeV}$}},
  \href{https://doi.org/10.1140/epjc/s10052-016-4293-4}{\emph{Eur. Phys. J. C}
  {\bfseries 76} (2016) 469}
  [\href{https://arxiv.org/abs/1603.01803}{{\ttfamily 1603.01803}}].

\bibitem{Sirunyan:2020oum}
{\scshape CMS} collaboration, \emph{{Measurements of the $W$ boson rapidity,
  helicity, double-differential cross sections, and charge asymmetry in $pp$
  collisions at $\sqrt {s} = 13~ \text{TeV}$}},
  \href{https://doi.org/10.1103/PhysRevD.102.092012}{\emph{Phys. Rev. D}
  {\bfseries 102} (2020) 092012}
  [\href{https://arxiv.org/abs/2008.04174}{{\ttfamily 2008.04174}}].

\bibitem{Aaboud:2016zkn}
{\scshape ATLAS} collaboration, \emph{{Search for new resonances in events with
  one lepton and missing transverse momentum in $pp$ collisions at $\sqrt{s} =
  13~ \text{TeV}$ with the ATLAS detector}},
  \href{https://doi.org/10.1016/j.physletb.2016.09.040}{\emph{Phys. Lett. B}
  {\bfseries 762} (2016) 334}
  [\href{https://arxiv.org/abs/1606.03977}{{\ttfamily 1606.03977}}].

\bibitem{Khachatryan:2016jww}
{\scshape CMS} collaboration, \emph{{Search for heavy gauge $W^{\prime}$ boson
  in events with an energetic lepton and large missing transverse momentum at
  $\sqrt{s} = 13~ \text{TeV}$}},
  \href{https://doi.org/10.1016/j.physletb.2017.04.043}{\emph{Phys. Lett. B}
  {\bfseries 770} (2017) 278}
  [\href{https://arxiv.org/abs/1612.09274}{{\ttfamily 1612.09274}}].

\bibitem{Aaboud:2017efa}
{\scshape ATLAS} collaboration, \emph{{Search for a new heavy gauge-boson
  resonance decaying into a lepton and missing transverse momentum in $36~
  \text{fb}^{-1}$ of $pp$ collisions at $\sqrt{s} = 13~ \text{TeV}$ with the
  ATLAS experiment}},
  \href{https://doi.org/10.1140/epjc/s10052-018-5877-y}{\emph{Eur. Phys. J. C}
  {\bfseries 78} (2018) 401}
  [\href{https://arxiv.org/abs/1706.04786}{{\ttfamily 1706.04786}}].

\bibitem{Aad:2019wvl}
{\scshape ATLAS} collaboration, \emph{{Search for a heavy charged boson in
  events with a charged lepton and missing transverse momentum from $pp$
  collisions at $\sqrt{s} = 13~ \text{TeV}$ with the ATLAS detector}},
  \href{https://doi.org/10.1103/PhysRevD.100.052013}{\emph{Phys. Rev. D}
  {\bfseries 100} (2019) 052013}
  [\href{https://arxiv.org/abs/1906.05609}{{\ttfamily 1906.05609}}].

\bibitem{Aaboud:2016cth}
{\scshape ATLAS} collaboration, \emph{{Search for high-mass new phenomena in
  the dilepton final state using proton-proton collisions at $\sqrt{s}=13~
  \text{TeV}$ with the ATLAS detector}},
  \href{https://doi.org/10.1016/j.physletb.2016.08.055}{\emph{Phys. Lett. B}
  {\bfseries 761} (2016) 372}
  [\href{https://arxiv.org/abs/1607.03669}{{\ttfamily 1607.03669}}].

\bibitem{Khachatryan:2016zqb}
{\scshape CMS} collaboration, \emph{{Search for narrow resonances in dilepton
  mass spectra in proton-proton collisions at $\sqrt{s} = 13~ \text{TeV}$ and
  combination with $8~ \text{TeV}$ data}},
  \href{https://doi.org/10.1016/j.physletb.2017.02.010}{\emph{Phys. Lett. B}
  {\bfseries 768} (2017) 57}
  [\href{https://arxiv.org/abs/1609.05391}{{\ttfamily 1609.05391}}].

\bibitem{Aad:2019fac}
{\scshape ATLAS} collaboration, \emph{{Search for high-mass dilepton resonances
  using $139~ \text{fb}^{-1}$ of $pp$ collision data collected at $\sqrt{s}=13~
  \text{TeV}$ with the ATLAS detector}},
  \href{https://doi.org/10.1016/j.physletb.2019.07.016}{\emph{Phys. Lett. B}
  {\bfseries 796} (2019) 68}
  [\href{https://arxiv.org/abs/1903.06248}{{\ttfamily 1903.06248}}].

\bibitem{Altarelli:1979ub}
G.~Altarelli, R.~K. Ellis and G.~Martinelli, \emph{{Large perturbative
  corrections to the Drell-Yan process in QCD}},
  \href{https://doi.org/10.1016/0550-3213(79)90116-0}{\emph{Nucl. Phys. B}
  {\bfseries 157} (1979) 461}.

\bibitem{Matsuura:1988sm}
T.~Matsuura, S.~C. van~der Marck and W.~L. van Neerven, \emph{{The calculation
  of the second order soft and virtual contributions to the Drell-Yan
  cross-section}},
  \href{https://doi.org/10.1016/0550-3213(89)90620-2}{\emph{Nucl. Phys. B}
  {\bfseries 319} (1989) 570}.

\bibitem{Hamberg:1990np}
R.~Hamberg, W.~L. van Neerven and T.~Matsuura, \emph{{A complete calculation of
  the order $\alpha_s^{2}$ correction to the Drell-Yan $K$-factor}},
  \href{https://doi.org/10.1016/0550-3213(91)90064-5}{\emph{Nucl. Phys. B}
  {\bfseries 359} (1991) 343}.

\bibitem{Harlander:2002wh}
R.~V. Harlander and W.~B. Kilgore, \emph{{Next-to-next-to-leading order Higgs
  production at hadron colliders}},
  \href{https://doi.org/10.1103/PhysRevLett.88.201801}{\emph{Phys. Rev. Lett.}
  {\bfseries 88} (2002) 201801}
  [\href{https://arxiv.org/abs/hep-ph/0201206}{{\ttfamily hep-ph/0201206}}].

\bibitem{Anastasiou:2003ds}
C.~Anastasiou, L.~Dixon, K.~Melnikov and F.~Petriello, \emph{{High-precision
  QCD at hadron colliders: Electroweak gauge boson rapidity distributions at
  next-to-next-to leading order}},
  \href{https://doi.org/10.1103/PhysRevD.69.094008}{\emph{Phys. Rev. D}
  {\bfseries 69} (2004) 094008}
  [\href{https://arxiv.org/abs/hep-ph/0312266}{{\ttfamily hep-ph/0312266}}].

\bibitem{Melnikov:2006di}
K.~Melnikov and F.~Petriello, \emph{{$W$ boson production cross section at the
  Large Hadron Collider with $\mathcal{O}(\alpha^2_s)$ corrections}},
  \href{https://doi.org/10.1103/PhysRevLett.96.231803}{\emph{Phys. Rev. Lett.}
  {\bfseries 96} (2006) 231803}
  [\href{https://arxiv.org/abs/hep-ph/0603182}{{\ttfamily hep-ph/0603182}}].

\bibitem{Melnikov:2006kv}
K.~Melnikov and F.~Petriello, \emph{{Electroweak gauge boson production at
  hadron colliders through $\mathcal{O}(\alpha_s^2)$}},
  \href{https://doi.org/10.1103/PhysRevD.74.114017}{\emph{Phys. Rev. D}
  {\bfseries 74} (2006) 114017}
  [\href{https://arxiv.org/abs/hep-ph/0609070}{{\ttfamily hep-ph/0609070}}].

\bibitem{Catani:2009sm}
S.~Catani, L.~Cieri, G.~Ferrera, D.~de~Florian and M.~Grazzini, \emph{{Vector
  boson production at hadron colliders: a fully exclusive QCD calculation at
  next-to-next-to-leading order}},
  \href{https://doi.org/10.1103/PhysRevLett.103.082001}{\emph{Phys. Rev. Lett.}
  {\bfseries 103} (2009) 082001}
  [\href{https://arxiv.org/abs/0903.2120}{{\ttfamily 0903.2120}}].

\bibitem{Catani:2010en}
S.~Catani, G.~Ferrera and M.~Grazzini, \emph{{$W$ boson production at hadron
  colliders: the lepton charge asymmetry in NNLO QCD}},
  \href{https://doi.org/10.1007/JHEP05(2010)006}{\emph{JHEP} {\bfseries 05}
  (2010) 006} [\href{https://arxiv.org/abs/1002.3115}{{\ttfamily 1002.3115}}].

\bibitem{Ahmed:2014cla}
T.~Ahmed, M.~Mahakhud, N.~Rana and V.~Ravindran, \emph{{Drell-Yan production at
  threshold to third order in QCD}},
  \href{https://doi.org/10.1103/PhysRevLett.113.112002}{\emph{Phys. Rev. Lett.}
  {\bfseries 113} (2014) 112002}
  [\href{https://arxiv.org/abs/1404.0366}{{\ttfamily 1404.0366}}].

\bibitem{Ahmed:2014uya}
T.~Ahmed, M.~K. Mandal, N.~Rana and V.~Ravindran, \emph{{Rapidity distributions
  in Drell-Yan and Higgs productions at threshold to third order in QCD}},
  \href{https://doi.org/10.1103/PhysRevLett.113.212003}{\emph{Phys. Rev. Lett.}
  {\bfseries 113} (2014) 212003}
  [\href{https://arxiv.org/abs/1404.6504}{{\ttfamily 1404.6504}}].

\bibitem{Duhr:2020seh}
C.~Duhr, F.~Dulat and B.~Mistlberger, \emph{{Drell-Yan cross section to third
  order in the strong coupling constant}},
  \href{https://doi.org/10.1103/PhysRevLett.125.172001}{\emph{Phys. Rev. Lett.}
  {\bfseries 125} (2020) 172001}
  [\href{https://arxiv.org/abs/2001.07717}{{\ttfamily 2001.07717}}].

\bibitem{Duhr:2020sdp}
C.~Duhr, F.~Dulat and B.~Mistlberger, \emph{{Charged current Drell-Yan
  production at $\text{N}^{3}\text{LO}$}},
  \href{https://doi.org/10.1007/JHEP11(2020)143}{\emph{JHEP} {\bfseries 11}
  (2020) 143} [\href{https://arxiv.org/abs/2007.13313}{{\ttfamily
  2007.13313}}].

\bibitem{Chen:2021vtu}
X.~Chen, T.~Gehrmann, N.~Glover, A.~Huss, T.-Z. Yang and H.~X. Zhu,
  \emph{{Dilepton rapidity distribution in Drell-Yan production to third rrder
  in QCD}}, \href{https://doi.org/10.1103/PhysRevLett.128.052001}{\emph{Phys.
  Rev. Lett.} {\bfseries 128} (2022) 052001}
  [\href{https://arxiv.org/abs/2107.09085}{{\ttfamily 2107.09085}}].

\bibitem{Baur:1997wa}
U.~Baur, S.~Keller and W.~K. Sakumoto, \emph{{QED radiative corrections to $Z$
  boson production and the forward-backward asymmetry at hadron colliders}},
  \href{https://doi.org/10.1103/PhysRevD.57.199}{\emph{Phys. Rev. D} {\bfseries
  57} (1998) 199} [\href{https://arxiv.org/abs/hep-ph/9707301}{{\ttfamily
  hep-ph/9707301}}].

\bibitem{Baur:2001ze}
U.~Baur, O.~Brein, W.~Hollik, C.~Schappacher and D.~Wackeroth,
  \emph{{Electroweak radiative corrections to neutral-current Drell-Yan
  processes at hadron colliders}},
  \href{https://doi.org/10.1103/PhysRevD.65.033007}{\emph{Phys. Rev. D}
  {\bfseries 65} (2002) 033007}
  [\href{https://arxiv.org/abs/hep-ph/0108274}{{\ttfamily hep-ph/0108274}}].

\bibitem{Zykunov:2006yb}
V.~A. Zykunov, \emph{{Radiative corrections to the Drell-Yan process at large
  dilepton invariant masses}},
  \href{https://doi.org/10.1134/S1063778806090109}{\emph{Phys. Atom. Nucl.}
  {\bfseries 69} (2006) 1522}.

\bibitem{Zykunov:2005tc}
V.~A. Zykunov, \emph{{Weak radiative corrections to the Drell-Yan process for
  large invariant mass of a dilepton pair}},
  \href{https://doi.org/10.1103/PhysRevD.75.073019}{\emph{Phys. Rev. D}
  {\bfseries 75} (2007) 073019}
  [\href{https://arxiv.org/abs/hep-ph/0509315}{{\ttfamily hep-ph/0509315}}].

\bibitem{CarloniCalame:2007cd}
C.~M. Carloni~Calame, G.~Montagna, O.~Nicrosini and A.~Vicini, \emph{{Precision
  electroweak calculation of the production of a high transverse-momentum
  lepton pair at hadron colliders}},
  \href{https://doi.org/10.1088/1126-6708/2007/10/109}{\emph{JHEP} {\bfseries
  10} (2007) 109} [\href{https://arxiv.org/abs/0710.1722}{{\ttfamily
  0710.1722}}].

\bibitem{Arbuzov:2007db}
A.~Arbuzov, D.~Bardin, S.~Bondarenko, P.~Christova, L.~Kalinovskaya, G.~Nanava
  et~al., \emph{{One-loop corrections to the Drell-Yan process in SANC (II) The
  neutral current case}},
  \href{https://doi.org/10.1140/epjc/s10052-008-0531-8}{\emph{Eur. Phys. J. C}
  {\bfseries 54} (2008) 451} [\href{https://arxiv.org/abs/0711.0625}{{\ttfamily
  0711.0625}}].

\bibitem{Dittmaier:2009cr}
S.~Dittmaier and M.~Huber, \emph{{Radiative corrections to the neutral-current
  Drell-Yan process in the Standard Model and its minimal supersymmetric
  extension}}, \href{https://doi.org/10.1007/JHEP01(2010)060}{\emph{JHEP}
  {\bfseries 01} (2010) 060} [\href{https://arxiv.org/abs/0911.2329}{{\ttfamily
  0911.2329}}].

\bibitem{Wackeroth:1996hz}
D.~Wackeroth and W.~Hollik, \emph{{Electroweak radiative corrections to
  resonant charged gauge boson production}},
  \href{https://doi.org/10.1103/PhysRevD.55.6788}{\emph{Phys. Rev. D}
  {\bfseries 55} (1997) 6788}
  [\href{https://arxiv.org/abs/hep-ph/9606398}{{\ttfamily hep-ph/9606398}}].

\bibitem{Baur:1998kt}
U.~Baur, S.~Keller and D.~Wackeroth, \emph{{Electroweak radiative corrections
  to $W$ boson production in hadronic collisions}},
  \href{https://doi.org/10.1103/PhysRevD.59.013002}{\emph{Phys. Rev. D}
  {\bfseries 59} (1999) 013002}
  [\href{https://arxiv.org/abs/hep-ph/9807417}{{\ttfamily hep-ph/9807417}}].

\bibitem{Dittmaier:2001ay}
S.~Dittmaier and M.~Kr\"amer, \emph{{Electroweak radiative corrections to
  $W$-boson production at hadron colliders}},
  \href{https://doi.org/10.1103/PhysRevD.65.073007}{\emph{Phys. Rev. D}
  {\bfseries 65} (2002) 073007}
  [\href{https://arxiv.org/abs/hep-ph/0109062}{{\ttfamily hep-ph/0109062}}].

\bibitem{Baur:2004ig}
U.~Baur and D.~Wackeroth, \emph{{Electroweak radiative corrections to
  $pp\hskip-7pt\hbox{$^{^{(\!-\!)}}$} \rightarrow W^{\pm} \rightarrow
  \ell^{\pm} \nu$ beyond the pole approximation}},
  \href{https://doi.org/10.1103/PhysRevD.70.073015}{\emph{Phys. Rev. D}
  {\bfseries 70} (2004) 073015}
  [\href{https://arxiv.org/abs/hep-ph/0405191}{{\ttfamily hep-ph/0405191}}].

\bibitem{Arbuzov:2005dd}
A.~Arbuzov, D.~Bardin, S.~Bondarenko, P.~Christova, L.~Kalinovskaya, G.~Nanava
  et~al., \emph{{One-loop corrections to the Drell-Yan process in SANC: the
  charged current case}},
  \href{https://doi.org/10.1140/epjc/s2006-02505-y}{\emph{Eur. Phys. J. C}
  {\bfseries 46} (2006) 407}
  [\href{https://arxiv.org/abs/hep-ph/0506110}{{\ttfamily hep-ph/0506110}}].

\bibitem{CarloniCalame:2006zq}
C.~M. Carloni~Calame, G.~Montagna, O.~Nicrosini and A.~Vicini, \emph{{Precision
  electroweak calculation of the charged current Drell-Yan process}},
  \href{https://doi.org/10.1088/1126-6708/2006/12/016}{\emph{JHEP} {\bfseries
  12} (2006) 016} [\href{https://arxiv.org/abs/hep-ph/0609170}{{\ttfamily
  hep-ph/0609170}}].

\bibitem{Brensing:2007qm}
S.~Brensing, S.~Dittmaier, M.~Kr\"amer and A.~M\"uck, \emph{{Radiative
  corrections to $W$-boson hadroproduction: Higher-order electroweak and
  supersymmetric effects}},
  \href{https://doi.org/10.1103/PhysRevD.77.073006}{\emph{Phys. Rev. D}
  {\bfseries 77} (2008) 073006}
  [\href{https://arxiv.org/abs/0710.3309}{{\ttfamily 0710.3309}}].

\bibitem{Alioli:2016fum}
S.~Alioli et~al., \emph{{Precision studies of observables in $pp \rightarrow W
  \rightarrow l\nu_l$ and $ pp \rightarrow \gamma,Z \rightarrow l^+l^-$
  processes at the LHC}},
  \href{https://doi.org/10.1140/epjc/s10052-017-4832-7}{\emph{Eur. Phys. J. C}
  {\bfseries 77} (2017) 280}
  [\href{https://arxiv.org/abs/1606.02330}{{\ttfamily 1606.02330}}].

\bibitem{Alioli:2015toa}
S.~Alioli, C.~W. Bauer, C.~Berggren, F.~J. Tackmann and J.~R. Walsh,
  \emph{{Drell-Yan production at $\text{NNLL}^{\prime}$+NNLO matched to parton
  showers}}, \href{https://doi.org/10.1103/PhysRevD.92.094020}{\emph{Phys. Rev.
  D} {\bfseries 92} (2015) 094020}
  [\href{https://arxiv.org/abs/1508.01475}{{\ttfamily 1508.01475}}].

\bibitem{Alioli:2016wqt}
S.~Alioli, C.~W. Bauer, S.~Guns and F.~J. Tackmann, \emph{{Underlying-event
  sensitive observables in Drell-Yan production using GENEVA}},
  \href{https://doi.org/10.1140/epjc/s10052-016-4458-1}{\emph{Eur. Phys. J. C}
  {\bfseries 76} (2016) 614}
  [\href{https://arxiv.org/abs/1605.07192}{{\ttfamily 1605.07192}}].

\bibitem{Camarda:2019zyx}
S.~Camarda, M.~Boonekamp, G.~Bozzi, S.~Catani, L.~Cieri, J.~Cuth et~al.,
  \emph{{DYTurbo: fast predictions for Drell-Yan processes}},
  \href{https://doi.org/10.1140/epjc/s10052-020-7757-5}{\emph{Eur. Phys. J. C}
  {\bfseries 80} (2020) 251}
  [\href{https://arxiv.org/abs/1910.07049}{{\ttfamily 1910.07049}}].

\bibitem{Boughezal:2016wmq}
R.~Boughezal, J.~M. Campbell, R.~K. Ellis, C.~Focke, W.~Giele, X.~Liu et~al.,
  \emph{{Color-singlet production at NNLO in MCFM}},
  \href{https://doi.org/10.1140/epjc/s10052-016-4558-y}{\emph{Eur. Phys. J. C}
  {\bfseries 77} (2017) 7} [\href{https://arxiv.org/abs/1605.08011}{{\ttfamily
  1605.08011}}].

\bibitem{Grazzini:2017mhc}
M.~Grazzini, S.~Kallweit and M.~Wiesemann, \emph{{Fully differential NNLO
  computations with MATRIX}},
  \href{https://doi.org/10.1140/epjc/s10052-018-5771-7}{\emph{Eur. Phys. J. C}
  {\bfseries 78} (2018) 537}
  [\href{https://arxiv.org/abs/1711.06631}{{\ttfamily 1711.06631}}].

\bibitem{Monni:2019whf}
P.~F. Monni, P.~Nason, E.~Re, M.~Wiesemann and G.~Zanderighi,
  \emph{{$\text{MiNNLO}_{\text{PS}}$: a new method to match NNLO QCD to parton
  showers}}, \href{https://doi.org/10.1007/JHEP05(2020)143}{\emph{JHEP}
  {\bfseries 05} (2020) 143}
  [\href{https://arxiv.org/abs/1908.06987}{{\ttfamily 1908.06987}}].

\bibitem{Alekhin:2021xcu}
S.~Alekhin, A.~Kardos, S.-O. Moch and Z.~Tr\'ocs\'anyi, \emph{{Precision
  studies for Drell-Yan processes at NNLO}},
  \href{https://doi.org/10.1140/epjc/s10052-021-09361-9}{\emph{Eur. Phys. J. C}
  {\bfseries 81} (2021) 573}
  [\href{https://arxiv.org/abs/2104.02400}{{\ttfamily 2104.02400}}].

\bibitem{deFlorian:2018wcj}
D.~de~Florian, M.~Der and I.~Fabre, \emph{{$\text{QCD}\oplus\text{QED}$ NNLO
  corrections to Drell-Yan production}},
  \href{https://doi.org/10.1103/PhysRevD.98.094008}{\emph{Phys. Rev. D}
  {\bfseries 98} (2018) 094008}
  [\href{https://arxiv.org/abs/1805.12214}{{\ttfamily 1805.12214}}].

\bibitem{Delto:2019ewv}
M.~Delto, M.~Jaquier, K.~Melnikov and R.~R\"ontsch, \emph{{Mixed
  $\text{QCD}\otimes\text{QED}$ corrections to on-shell $Z$ boson production at
  the LHC}}, \href{https://doi.org/10.1007/JHEP01(2020)043}{\emph{JHEP}
  {\bfseries 01} (2020) 043}
  [\href{https://arxiv.org/abs/1909.08428}{{\ttfamily 1909.08428}}].

\bibitem{Cieri:2020ikq}
L.~Cieri, D.~de~Florian, M.~Der and J.~Mazzitelli, \emph{{Mixed
  $\text{QCD}\otimes\text{QED}$ corrections to exclusive Drell Yan production
  using the $q_T$-subtraction method}},
  \href{https://doi.org/10.1007/JHEP09(2020)155}{\emph{JHEP} {\bfseries 09}
  (2020) 155} [\href{https://arxiv.org/abs/2005.01315}{{\ttfamily
  2005.01315}}].

\bibitem{Bonciani:2019nuy}
R.~Bonciani, F.~Buccioni, N.~Rana, I.~Triscari and A.~Vicini, \emph{{NNLO
  $\text{QCD}\times\text{EW}$ corrections to $Z$ production in the $q\bar{q}$
  channel}}, \href{https://doi.org/10.1103/PhysRevD.101.031301}{\emph{Phys.
  Rev. D} {\bfseries 101} (2020) 031301}
  [\href{https://arxiv.org/abs/1911.06200}{{\ttfamily 1911.06200}}].

\bibitem{Buccioni:2020cfi}
F.~Buccioni, F.~Caola, M.~Delto, M.~Jaquier, K.~Melnikov and R.~R\"ontsch,
  \emph{{Mixed QCD-electroweak corrections to on-shell $Z$ production at the
  LHC}}, \href{https://doi.org/10.1016/j.physletb.2020.135969}{\emph{Phys.
  Lett. B} {\bfseries 811} (2020) 135969}
  [\href{https://arxiv.org/abs/2005.10221}{{\ttfamily 2005.10221}}].

\bibitem{Bonciani:2020tvf}
R.~Bonciani, F.~Buccioni, N.~Rana and A.~Vicini, \emph{{Next-to-next-to-leading
  order mixed QCD-electroweak corrections to on-shell $Z$ production}},
  \href{https://doi.org/10.1103/PhysRevLett.125.232004}{\emph{Phys. Rev. Lett.}
  {\bfseries 125} (2020) 232004}
  [\href{https://arxiv.org/abs/2007.06518}{{\ttfamily 2007.06518}}].

\bibitem{Behring:2020cqi}
A.~Behring, F.~Buccioni, F.~Caola, M.~Delto, M.~Jaquier, K.~Melnikov et~al.,
  \emph{{Mixed QCD-electroweak corrections to $W$-boson production in hadron
  collisions}}, \href{https://doi.org/10.1103/PhysRevD.103.013008}{\emph{Phys.
  Rev. D} {\bfseries 103} (2021) 013008}
  [\href{https://arxiv.org/abs/2009.10386}{{\ttfamily 2009.10386}}].

\bibitem{Dittmaier:2020vra}
S.~Dittmaier, T.~Schmidt and J.~Schwarz, \emph{{Mixed NNLO
  $\text{QCD}\times\text{electroweak}$ corrections of $\mathcal{O}(N_f \alpha_s
  \alpha)$ to single-$W/Z$ production at the LHC}},
  \href{https://doi.org/10.1007/JHEP12(2020)201}{\emph{JHEP} {\bfseries 12}
  (2020) 201} [\href{https://arxiv.org/abs/2009.02229}{{\ttfamily
  2009.02229}}].

\bibitem{Dittmaier:2014qza}
S.~Dittmaier, A.~Huss and C.~Schwinn, \emph{{Mixed QCD-electroweak
  $\mathcal{O}(\alpha_s\alpha)$ corrections to Drell-Yan processes in the
  resonance region: Pole approximation and non-factorizable corrections}},
  \href{https://doi.org/10.1016/j.nuclphysb.2014.05.027}{\emph{Nucl. Phys. B}
  {\bfseries 885} (2014) 318}
  [\href{https://arxiv.org/abs/1403.3216}{{\ttfamily 1403.3216}}].

\bibitem{Dittmaier:2015rxo}
S.~Dittmaier, A.~Huss and C.~Schwinn, \emph{{Dominant mixed QCD-electroweak
  $\mathcal{O}(\alpha_s\alpha)$ corrections to Drell-Yan processes in the
  resonance region}},
  \href{https://doi.org/10.1016/j.nuclphysb.2016.01.006}{\emph{Nucl. Phys. B}
  {\bfseries 904} (2016) 216}
  [\href{https://arxiv.org/abs/1511.08016}{{\ttfamily 1511.08016}}].

\bibitem{Bonciani:2016ypc}
R.~Bonciani, S.~Di~Vita, P.~Mastrolia and U.~Schubert, \emph{{Two-loop master
  integrals for the mixed EW-QCD virtual corrections to Drell-Yan scattering}},
  \href{https://doi.org/10.1007/JHEP09(2016)091}{\emph{JHEP} {\bfseries 09}
  (2016) 091} [\href{https://arxiv.org/abs/1604.08581}{{\ttfamily
  1604.08581}}].

\bibitem{vonManteuffel:2017myy}
A.~von Manteuffel and R.~M. Schabinger, \emph{{Numerical multi-loop
  calculations via finite integrals and one-mass EW-QCD Drell-Yan master
  integrals}}, \href{https://doi.org/10.1007/JHEP04(2017)129}{\emph{JHEP}
  {\bfseries 04} (2017) 129}
  [\href{https://arxiv.org/abs/1701.06583}{{\ttfamily 1701.06583}}].

\bibitem{Heller:2019gkq}
M.~Heller, A.~von Manteuffel and R.~M. Schabinger, \emph{{Multiple
  polylogarithms with algebraic arguments and the two-loop EW-QCD Drell-Yan
  master integrals}},
  \href{https://doi.org/10.1103/PhysRevD.102.016025}{\emph{Phys. Rev. D}
  {\bfseries 102} (2020) 016025}
  [\href{https://arxiv.org/abs/1907.00491}{{\ttfamily 1907.00491}}].

\bibitem{Heller:2020owb}
M.~Heller, A.~von Manteuffel, R.~M. Schabinger and H.~Spiesberger, \emph{{Mixed
  EW-QCD two-loop amplitudes for $q\bar{q} \rightarrow \ell^+\ell^-$ and
  $\gamma_5$ scheme independence of multi-loop corrections}},
  \href{https://doi.org/10.1007/JHEP05(2021)213}{\emph{JHEP} {\bfseries 05}
  (2021) 213} [\href{https://arxiv.org/abs/2012.05918}{{\ttfamily
  2012.05918}}].

\bibitem{Bonciani:2021zzf}
R.~Bonciani, L.~Buonocore, M.~Grazzini, S.~Kallweit, N.~Rana, F.~Tramontano
  et~al., \emph{{Mixed strong-electroweak corrections to the Drell-Yan
  process}}, \href{https://doi.org/10.1103/PhysRevLett.128.012002}{\emph{Phys.
  Rev. Lett.} {\bfseries 128} (2022) 012002}
  [\href{https://arxiv.org/abs/2106.11953}{{\ttfamily 2106.11953}}].

\bibitem{Kinoshita:1962ur}
T.~Kinoshita, \emph{{Mass singularities of Feynman amplitudes}},
  \href{https://doi.org/10.1063/1.1724268}{\emph{J. Math. Phys.} {\bfseries 3}
  (1962) 650}.

\bibitem{Lee:1964is}
T.~D. Lee and M.~Nauenberg, \emph{{Degenerate systems and mass singularities}},
  \href{https://doi.org/10.1103/PhysRev.133.B1549}{\emph{Phys. Rev.} {\bfseries
  133} (1964) B1549}.

\bibitem{Hasan:2020vwn}
S.~M. Hasan and U.~Schubert, \emph{{Master integrals for the mixed QCD-QED
  corrections to the Drell-Yan production of a massive lepton pair}},
  \href{https://doi.org/10.1007/JHEP11(2020)107}{\emph{JHEP} {\bfseries 11}
  (2020) 107} [\href{https://arxiv.org/abs/2004.14908}{{\ttfamily
  2004.14908}}].

\bibitem{Binosi:2008ig}
D.~Binosi, J.~Collins, C.~Kaufhold and L.~Theussl, \emph{{JaxoDraw: A Graphical
  user interface for drawing Feynman diagrams. Version 2.0 release notes}},
  \href{https://doi.org/10.1016/j.cpc.2009.02.020}{\emph{Comput. Phys. Commun.}
  {\bfseries 180} (2009) 1709}
  [\href{https://arxiv.org/abs/0811.4113}{{\ttfamily 0811.4113}}].

\bibitem{Chetyrkin:1981qh}
K.~G. Chetyrkin and F.~V. Tkachov, \emph{{Integration by parts: the algorithm
  to calculate $\beta$-functions in $4$ Loops}},
  \href{https://doi.org/10.1016/0550-3213(81)90199-1}{\emph{Nucl. Phys. B}
  {\bfseries 192} (1981) 159}.

\bibitem{Kotikov:1990kg}
A.~V. Kotikov, \emph{{Differential equations method. New technique for massive
  Feynman diagram calculation}},
  \href{https://doi.org/10.1016/0370-2693(91)90413-K}{\emph{Phys. Lett. B}
  {\bfseries 254} (1991) 158}.

\bibitem{Remiddi:1997ny}
E.~Remiddi, \emph{{Differential equations for Feynman graph amplitudes}},
  \href{https://doi.org/10.1007/BF03185566}{\emph{Nuovo Cim. A} {\bfseries 110}
  (1997) 1435} [\href{https://arxiv.org/abs/hep-th/9711188}{{\ttfamily
  hep-th/9711188}}].

\bibitem{Maierhoefer:2017hyi}
P.~Maierh\"ofer, J.~Usovitsch and P.~Uwer, \emph{{Kira\textemdash{}A Feynman
  integral reduction program}},
  \href{https://doi.org/10.1016/j.cpc.2018.04.012}{\emph{Comput. Phys. Commun.}
  {\bfseries 230} (2018) 99}
  [\href{https://arxiv.org/abs/1705.05610}{{\ttfamily 1705.05610}}].

\bibitem{Klappert:2020nbg}
J.~Klappert, F.~Lange, P.~Maierh\"ofer and J.~Usovitsch, \emph{{Integral
  reduction with Kira 2.0 and finite field methods}},
  \href{https://doi.org/10.1016/j.cpc.2021.108024}{\emph{Comput. Phys. Commun.}
  {\bfseries 266} (2021) 108024}
  [\href{https://arxiv.org/abs/2008.06494}{{\ttfamily 2008.06494}}].

\bibitem{Laporta:2000dsw}
S.~Laporta, \emph{{High precision calculation of multiloop Feynman integrals by
  difference equations}},
  \href{https://doi.org/10.1142/S0217751X00002159}{\emph{Int. J. Mod. Phys. A}
  {\bfseries 15} (2000) 5087}
  [\href{https://arxiv.org/abs/hep-ph/0102033}{{\ttfamily hep-ph/0102033}}].

\bibitem{Lee:2012cn}
R.~N. Lee, \emph{{Presenting LiteRed: a tool for the Loop InTEgrals
  REDuction}},  \href{https://arxiv.org/abs/1212.2685}{{\ttfamily 1212.2685}}.

\bibitem{Lee:2013mka}
R.~N. Lee, \emph{{LiteRed 1.4: a powerful tool for reduction of multiloop
  integrals}}, \href{https://doi.org/10.1088/1742-6596/523/1/012059}{\emph{J.
  Phys. Conf. Ser.} {\bfseries 523} (2014) 012059}
  [\href{https://arxiv.org/abs/1310.1145}{{\ttfamily 1310.1145}}].

\bibitem{Henn:2013pwa}
J.~M. Henn, \emph{{Multiloop integrals in dimensional regularization made
  simple}}, \href{https://doi.org/10.1103/PhysRevLett.110.251601}{\emph{Phys.
  Rev. Lett.} {\bfseries 110} (2013) 251601}
  [\href{https://arxiv.org/abs/1304.1806}{{\ttfamily 1304.1806}}].

\bibitem{Goncharov:2001iea}
A.~B. Goncharov, \emph{{Multiple polylogarithms and mixed Tate motives}},
  \href{https://arxiv.org/abs/math/0103059}{{\ttfamily math/0103059}}.

\bibitem{Goncharov:1998kja}
A.~B. Goncharov, \emph{{Multiple polylogarithms, cyclotomy and modular
  complexes}}, \href{https://doi.org/10.4310/MRL.1998.v5.n4.a7}{\emph{Math.
  Res. Lett.} {\bfseries 5} (1998) 497}
  [\href{https://arxiv.org/abs/1105.2076}{{\ttfamily 1105.2076}}].

\bibitem{Maitre:2005uu}
D.~Ma\^{i}tre, \emph{{HPL, a Mathematica implementation of the harmonic
  polylogarithms}},
  \href{https://doi.org/10.1016/j.cpc.2005.10.008}{\emph{Comput. Phys. Commun.}
  {\bfseries 174} (2006) 222}
  [\href{https://arxiv.org/abs/hep-ph/0507152}{{\ttfamily hep-ph/0507152}}].

\bibitem{Maitre:2007kp}
D.~Ma\^{i}tre, \emph{{Extension of HPL to complex arguments}},
  \href{https://doi.org/10.1016/j.cpc.2011.11.015}{\emph{Comput. Phys. Commun.}
  {\bfseries 183} (2012) 846}
  [\href{https://arxiv.org/abs/hep-ph/0703052}{{\ttfamily hep-ph/0703052}}].

\bibitem{Duhr:2019tlz}
C.~Duhr and F.~Dulat, \emph{{PolyLogTools \textemdash{} polylogs for the
  masses}}, \href{https://doi.org/10.1007/JHEP08(2019)135}{\emph{JHEP}
  {\bfseries 08} (2019) 135}
  [\href{https://arxiv.org/abs/1904.07279}{{\ttfamily 1904.07279}}].

\bibitem{Bauer:2000cp}
C.~W. Bauer, A.~Frink and R.~Kreckel, \emph{{Introduction to the GiNaC
  framework for symbolic computation within the C++ programming language}},
  \href{https://doi.org/10.1006/jsco.2001.0494}{\emph{J. Symb. Comput.}
  {\bfseries 33} (2002) 1} [\href{https://arxiv.org/abs/cs/0004015}{{\ttfamily
  cs/0004015}}].

\bibitem{Vollinga:2004sn}
J.~Vollinga and S.~Weinzierl, \emph{{Numerical evaluation of multiple
  polylogarithms}},
  \href{https://doi.org/10.1016/j.cpc.2004.12.009}{\emph{Comput. Phys. Commun.}
  {\bfseries 167} (2005) 177}
  [\href{https://arxiv.org/abs/hep-ph/0410259}{{\ttfamily hep-ph/0410259}}].

\bibitem{Mastrolia:2017pfy}
P.~Mastrolia, M.~Passera, A.~Primo and U.~Schubert, \emph{{Master integrals for
  the NNLO virtual corrections to $\mu e$ scattering in QED: the planar
  graphs}}, \href{https://doi.org/10.1007/JHEP11(2017)198}{\emph{JHEP}
  {\bfseries 11} (2017) 198}
  [\href{https://arxiv.org/abs/1709.07435}{{\ttfamily 1709.07435}}].

\bibitem{Borowka:2017idc}
S.~Borowka, G.~Heinrich, S.~Jahn, S.~P. Jones, M.~Kerner, J.~Schlenk et~al.,
  \emph{{pySecDec: A toolbox for the numerical evaluation of multi-scale
  integrals}}, \href{https://doi.org/10.1016/j.cpc.2017.09.015}{\emph{Comput.
  Phys. Commun.} {\bfseries 222} (2018) 313}
  [\href{https://arxiv.org/abs/1703.09692}{{\ttfamily 1703.09692}}].

\bibitem{Borowka:2018goh}
S.~Borowka, G.~Heinrich, S.~Jahn, S.~P. Jones, M.~Kerner and J.~Schlenk,
  \emph{{A GPU compatible quasi-Monte Carlo integrator interfaced to
  pySecDec}}, \href{https://doi.org/10.1016/j.cpc.2019.02.015}{\emph{Comput.
  Phys. Commun.} {\bfseries 240} (2019) 120}
  [\href{https://arxiv.org/abs/1811.11720}{{\ttfamily 1811.11720}}].

\bibitem{Gehrmann:2014bfa}
T.~Gehrmann, A.~von Manteuffel, L.~Tancredi and E.~Weihs, \emph{{The two-loop
  master integrals for $q\bar{q} \rightarrow VV$}},
  \href{https://doi.org/10.1007/JHEP06(2014)032}{\emph{JHEP} {\bfseries 06}
  (2014) 032} [\href{https://arxiv.org/abs/1404.4853}{{\ttfamily 1404.4853}}].

\bibitem{Panzer:2014caa}
E.~Panzer, \emph{{Algorithms for the symbolic integration of hyperlogarithms
  with applications to Feynman integrals}},
  \href{https://doi.org/10.1016/j.cpc.2014.10.019}{\emph{Comput. Phys. Commun.}
  {\bfseries 188} (2015) 148}
  [\href{https://arxiv.org/abs/1403.3385}{{\ttfamily 1403.3385}}].

\bibitem{Pak:2010pt}
A.~V. Pak and A.~V. Smirnov, \emph{{Geometric approach to asymptotic expansion
  of Feynman integrals}},
  \href{https://doi.org/10.1140/epjc/s10052-011-1626-1}{\emph{Eur. Phys. J. C}
  {\bfseries 71} (2011) 1626}
  [\href{https://arxiv.org/abs/1011.4863}{{\ttfamily 1011.4863}}].

\bibitem{Jantzen:2012mw}
B.~Jantzen, A.~V. Smirnov and V.~A. Smirnov, \emph{{Expansion by regions:
  revealing potential and Glauber regions automatically}},
  \href{https://doi.org/10.1140/epjc/s10052-012-2139-2}{\emph{Eur. Phys. J. C}
  {\bfseries 72} (2012) 2139}
  [\href{https://arxiv.org/abs/1206.0546}{{\ttfamily 1206.0546}}].

\bibitem{Smirnov:2015mct}
A.~V. Smirnov, \emph{{FIESTA4: Optimized Feynman integral calculations with GPU
  support}}, \href{https://doi.org/10.1016/j.cpc.2016.03.013}{\emph{Comput.
  Phys. Commun.} {\bfseries 204} (2016) 189}
  [\href{https://arxiv.org/abs/1511.03614}{{\ttfamily 1511.03614}}].

\bibitem{Smirnov:2021rhf}
A.~V. Smirnov, N.~D. Shapurov and L.~I. Vysotsky, \emph{{FIESTA5: numerical
  high-performance Feynman integral evaluation}},
  \href{https://arxiv.org/abs/2110.11660}{{\ttfamily 2110.11660}}.

\end{thebibliography}\endgroup

\end{document}